\documentclass[11pt,a4paper]{article}
\usepackage{jheppub}

\usepackage{graphicx}% Include figure files
\usepackage{dcolumn}% Align table columns on decimal point
\usepackage{bm}% bold math
\usepackage{subfig}
\global\long\def\mC{\mathcal{C}}%
\global\long\def\mG{\mathcal{G}}%
\global\long\def\mJ{\mathcal{J}}%
\global\long\def\mK{\mathcal{K}}%
\global\long\def\mO{\mathcal{O}}%
\global\long\def\mT{\mathcal{T}}%
\global\long\def\mU{\mathcal{U}}%
\global\long\def\e{\epsilon}%
\global\long\def\ra{\rightarrow}%
\global\long\def\avg#1{\left\langle #1\right\rangle }%
\global\long\def\f#1#2{\frac{#1}{#2}}%
\global\long\def\del{\partial}%
\global\long\def\t{\theta}%
\global\long\def\a{\alpha}%
\global\long\def\b{\beta}%
\global\long\def\g{\gamma}%
\global\long\def\s{\sigma}%
\global\long\def\r{\rho}%
\global\long\def\d{\delta}%
\global\long\def\ket#1{\left\langle #1\right|}%
\global\long\def\bra#1{\left|#1\right\rangle }%
\global\long\def\N{\mathbb{N}}%
\global\long\def\I{\mathbb{I}}%
\global\long\def\Z{\mathbb{Z}}%
\global\long\def\R{\mathbb{R}}%
\global\long\def\p{\varphi}%
\global\long\def\T{\text{T}}%
\global\long\def\w{\omega}%
\global\long\def\D{\Delta}%
\global\long\def\S{\Sigma}%
\global\long\def\app{\approx}%
\global\long\def\sgn{\text{sgn}}%

\newcommand{\be}{\begin{equation}}
\newcommand{\ee}{\end{equation}}
\def\nn{\nonumber}
\def\max{\text{max}}
\def\min{\text{min}}

\begin{document}

\title{A Traversable Wormhole Teleportation Protocol in the SYK Model}

\author[a]{Ping Gao}
\author[b]{and Daniel Louis Jafferis}
 \affiliation[a]{Center for Theoretical Physics, Massachusetts Institute of Technology,\\ Cambridge, MA 02139, USA}
\affiliation[b]{Center for the Fundamental Laws of Nature, Harvard University,\\ 
Cambridge, MA 02138, USA}

\emailAdd{pgao@mit.edu}
\emailAdd{jafferis@g.harvard.edu}

\abstract{
In this paper, we propose a concrete teleportation protocol in the SYK model based on a particle traversing a wormhole. The required operations for the communication, and insertion and extraction of the qubit, are all simple operators in terms of the basic qubits. We determine the effectiveness of this protocol, and find a version achieves almost perfect fidelity. Many features of semiclassical traversable wormholes are manifested in this setup.
}

%\keywords{Traversable wormholes, Teleportation, SYK model}%Use showkeys class option if keyword
                              %display desired
\maketitle

%\tableofcontents

\section{Introduction}

Recent constructions of traversable wormholes \cite{gao2017traversable,maldacena2017diving} provide  a  causal probe of the  ER=EPR \cite{maldacena2013cool} relation between entanglement and geometry. A pair of black holes in thermofield double state have their interiors connected via a Einstein-Rosen bridge behind the horizon. The Einstein-Rosen bridge is non-traversable so signals cannot get through the wormhole although the black hole interiors are spatially connected. The basic mechanism found in \cite{gao2017traversable} is that the gravitational backreaction to quantum effects induced by generic couplings between the exterior regions of the pair of black holes can render the wormhole traversable. One of the most interesting consequences is that a region of connected spacetime which would have been cloaked behind the horizon becomes visible to the external boundaries. Many recent works have studied traversable wormholes in more general settings, {\it e.g.} \cite{Maldacena:2020sxe, Maldacena:2018gjk, Fallows:2020ugr, Emparan:2020ldj, Balushi:2020lyc}.

It was conjectured by \cite{gao2017traversable,maldacena2017diving} that sending quantum information through such a wormhole is the  gravitational description of quantum teleportation in the dual many-body system. Let us denote the two black holes by $l$ and $r$, which are initially in the thermofield double state. The physical picture behind this teleportation is quite simple: the qubit that is teleported from $l$ to $r$ is carried by a  particle  that travels through the dual wormhole.

For a qubit thrown into one side of the system at a sufficiently early time, one just needs to wait long enough (of order the  scrambling time) so that the qubit is sufficiently mixed with the black hole microstates, then apply a simple coupling between $l$ and $r$, and finally wait for another scrambling time for the qubit to appear on the other side.

Many subsequent works developed this picture in more detail. 
In \cite{susskind2018teleportation}, a scrambling process combined with teleportation protocol was compared with the traversable wormhole protocol. Their protocol requires precursor operators with a much higher complexity than the simple coupling between $l$ and $r$. In \cite{bao2018traversable}, the traversable wormhole was understood as a quantum channel. In \cite{yoshida2017efficient}, a nontrivial protocol was proposed to decode a qubit that has been highly scrambled in a black hole background. However,  not all of the  features of the protocol suggested by traversable wormholes are manifested in these versions, particularly that fact that only simple operators are required. \footnote{A very recent work \cite{adambrown2019quantumgra} proposed a protocol of teleportation by size of operators, which in that sense is similar to traversable wormholes.} Moreover, as we will discuss further below, other effects can lead to teleportation associated to fully scrambled (essentially random unitarity) dynamics, one sign of which is time inversion of the transmitted quantum information. 

In this paper, we propose a precise teleportation protocol using $N$ qubit fermionic systems that correspond to gravity in a nearly AdS$_2$ black hole spacetime. We will determine the fidelity of teleportation by computing the mutual information between a reference qubit that is initially in a Bell state with the injected qubit on the left and a final register qubit on the right. We find an optimal method for encoding the qubit in terms of the basic fermions that leads to perfect teleportation for large $N$ in a certain limit on the parameters describing the dynamics of fermionic system. More generally, we also investigate the separability of the final state when the mutual information is sub-maximal, to determine when the information transfer is truly quantum. 

We examine this in detail in the SYK model \cite{sachdev1993gapless,kitaev2015simple} with $q$-point random interactions \cite{maldacena2016remarks}, which is solvable in large $N$ limit. % and dual to AdS$_2$.  
 Our protocol 
invovles applying SWAP operations constructed out of simple operators of the SYK model to insert the qubit into the black hole from the left side and  extract it  on the right side, as well as a communication channel between $l$ and $r$ that involves a coupling of only the basic fermions, of the type used in \cite{maldacena2017diving, maldacena2018eternal} (see Fig. \ref{fig:protocol}). Recent works studying traversable wormholes using the SYK model can been found in {\it e.g.} \cite{Numasawa:2020sty, Sorokhaibam:2020ilg, Chen:2019qqe, Lensky:2020fqf, Maldacena:2019ufo, Plugge:2020wgc, Qi:2020ian}.

One can construct Dirac fermions out of pairs of the basic Majoranas, and the teleported qubits are encoded in those Dirac fermion occupation numbers. It is particularly interesting to examine situations which have a clear dual description in terms of a particle physically propagating through the wormhole. If our qubit was encoded in the occupation number of a bulk fermion mode, then at low temperatures, it would be distinguishable from the thermal atmosphere. Note that at high temperatures, there is a significant probability that a given mode is already occupied in the Hartle-Hawking state itself. %\PG{this argument holds only when there is a chemical potential for occupation number. If not, like SYK, fermion number occupation is far from a good quantum number to distinguish energy eigenstates. I think pointing this out would explain the statement in the paragraph below.} \DLJ{I think that for the bulk fermion, it must be unlikely to be occupied in a low temperature black hole. This seems obvious in the bulk - adding a bulk fermion must increase the energy by its rest mass. The point is that the bulk fermion occupation number (far from the horizon) is $\int dt dt' f(t)^* f(t') \psi^\dag(t) \psi(t')$, and this will be small in the tfd at low temperature because of the time dependence of the 2 point function. If that's not true, how is it consistent with thinking about a bulk particle at all.}

The bulk description of the Majorana SYK model is slightly more complicated at the boundary, and the fundamental fermions have occupation number $\langle \hat N \rangle = \frac{1}{2}$ in the thermal state at any temperature. Nevertheless, the correlation function between the number operators in the left and right systems are small at low temperature, and the desired behavior can still be seen. In particular, after SWAP the qubit into SYK, the occupation number of fundamental fermions will change according to the qubit. %\PG{I think the picture for fermions carrying information is very tricky for us. For a specific state $\bra{Q}=a\bra{0}+b\bra{1}$, after SWAP, one can show that the fermion number right after the SWAP is $|b|^2 \neq 1/2$ (at the SWAPed site). It is the information itself that determines the change of occupation number. Generally, as the Hamiltonian is not commutative with occupation number, after SWAP, the occupation number will change along time. I found the following result. Suppose SWAP at $t=0$, and the occupation number at $t$ is $|a|^2(1+A+B)/2+|b|^2(1-A-B)/2$, where $A=\avg{\psi_1 \psi_2(t)\psi_1(t)\psi_2}-\avg{\psi_2\psi_2(t)\psi_1(t)\psi_1}$, and $B=\avg{\psi_2 \psi_1 \psi_2(t)\psi_1(t)}+\avg{\psi_2(t)\psi_1(t)\psi_2\psi_1}$. Note that $A$ is OTOC and $B$ is TOC. We can estimate that at early time $A=B=-1/2$, then occupation is $|b|^2$; at late time $A\ra 0$ and $B=-1/2$, then occupation is $3|b|^2/4+|a|^2/4$. Hence, I would say thinking about our protocol as encoding the qubit into occupation of a specific fermion is not quite accurate. Moreover, when the particle moves into the deep interior and the qubit is scrambled, the operator probing its existence in the bulk is no longer the occupation number on the boundary. Nevertheless, we need to emphasize that the qubit is not encoded into an internal space (say spin) of a  particle, which travels through the traversable wormhole just like a spaceship carrying a passenger. In that case, the information is stored in a factorized Hilbert space and is not scrambled. It is important that the information is encoded in the existence of particle (in the wavefunction sense), namely occupation number, and we cannot simply say this particle travels through just like a classical one. I think this problem is related to how a teleportee feels in traversable wormhole, which needs further investigation.}

We find the associated solution in the classical large $N$ limit in terms of the collective bilocal variables $G$ and $\Sigma$. To solve the classical equations we further specialize to the large $q$ limit where they simplify. Even at low temperatures, our analysis goes beyond the leading Schwarzian approximation. 

The effect of the instantaneous interaction required for teleportation can be described exactly in terms of its action on the basic Majoranas. This results in a non-trivial transformation of the bilocal field $G$ that we determine. Thus in the large $N$ limit, the problem reduces to finding a solution with that twist boundary condition across the moment of interaction.

Features of the full $G$-$\Sigma$ solution beyond the Schwarzian mode are important in understanding teleportation. In particular, in the Schwarzian approximation, the leading order $1/N$ classical solution for the collective reparametrization field due to the interaction would lead to a pole in the response measured by the occupation number on the right, whereas Fermi statistics demands that this number is bounded by 1. We will find a consistent result from classical $G$-$\Sigma$ solution, which allows us to determine the fidelity of the teleportation protocol. 

An exact bulk dual of the SYK model is not yet known, but we will often refer to the gravity description for intuition and to interpret our results. At low temperatures in the large $N$ limit, the correlation functions of fermions in SYK agree with free fields in the two dimensional bulk spacetime, corrected at leading order in $1/N$ by exchanges of the Schwarzian soft mode. This approximation, which is equivalent to bulk fermions minimally coupled to Jackiw-Teitelboim gravity, suffices to match the physics of traversable wormholes in the appropriate regime. 

On the gravity side, stringy corrections lead to inelastic effects in the out of time order correlators that describe quantum chaos \cite{Shenker:2014cwa, Gu:2018jsv}, and we will analogously refer to such features in SYK in that way. The $G$-$\Sigma$ boundary bi-local variables, however, have not yet been recast in a manner that looks like a bulk string field theory. 

%We show the effectiveness of this protocol in SYK model \cite{sachdev1993gapless,kitaev2015simple} which is dual to near $AdS_2$ black holes. 

We find that many features of semiclassical traversable wormholes are manifested in this setup. In particular, the coupling between $l$ and $r$ can be rather general while maintaining the fidelity of the protocol. Moreover, if we insert a sequence of qubits into the system in an appropriate time window, these qubits will come out with the same time ordering, as indicated by the smooth semiclassical geometry of a traversable wormhole. 

This should be contrasted with what occurs in what one might call the fully scrambled regime, when the evolution $e^{i H t}$ is effectively equivalent to a random unitary. Signals can still reach the right, however they appear in reverse time order. We check that with our protocol some quantum information can be sent in this case only at very high temperatures, although the fidelity is submaximal. 

At low temperatures in the gravity description, the fully scrambled regime corresponds to a particle sent so early into the black hole that it has a destructive ultra high energy collision with the negative energy squeezed state generated by the interaction protocol and  it does not smoothly pass through the wormhole. Nevertheless signals can be detected on the right due to a quantum interference effect \cite{maldacena2017diving, gao2018regenesis}. Our protocol involves encoding a qubit into one of the fermions, rather than simple signaling as measured by a left-right causal propagator. It would be interesting to understand if there is a bulk description of the partially successful teleportation we find at very high temperatures in the fully scrambled regime. 

We also find another regime in which qubits arrive on the right in inverse time order. This appears at any temperature when the qubit is injected too late. In that case, the traversing trajectory will be very close to ultimate horizon of the traverable wormhole, and it is likely that quantum fluctuations of the geometry, or stringy effects related to the AdS boundary in whatever is the exact dual of the SYK model are important. We leave further exploration of that effect in the bulk description to future work.

An important feature of the traversable wormhole teleportation protocol is that the qubit travels through the deep interior of the spacetime, and thus it is insensitive to single boundary simple unitary operations after the qubit is sufficiently scrambled. Therefore this is teleportation of a highly error protected qubit.

The structure of this paper is as follows. In Sec. \ref{sec:a-simple}, we introduce the protocol and describe its detailed realization  in the SYK model. To assess the effectiveness of the protocol, we need to calculate various correlation functions, which we do in Sec. \ref{sec:solve},  in the large $q$ limit. In Sec. \ref{sec:effect}, we demonstrate the effectiveness of the protocol. In Sec. \ref{sec:discuss} we discuss four aspects of the protocol: the fully scrambled interference regime, more general choices of coupling, teleportation of multiple qubits, and whether any quantum information is transmitted in cases with submaximal fidelity. We conclude in Sec. \ref{sec:conclusion}.  Appendix \ref{app} contains further details of the solution of the key correlation function.

\section{A simple wormhole teleportation protocol}\label{sec:a-simple}

\subsection{General setup} \label{sec:setup}

\begin{figure}
\centering
\includegraphics[width=5cm]{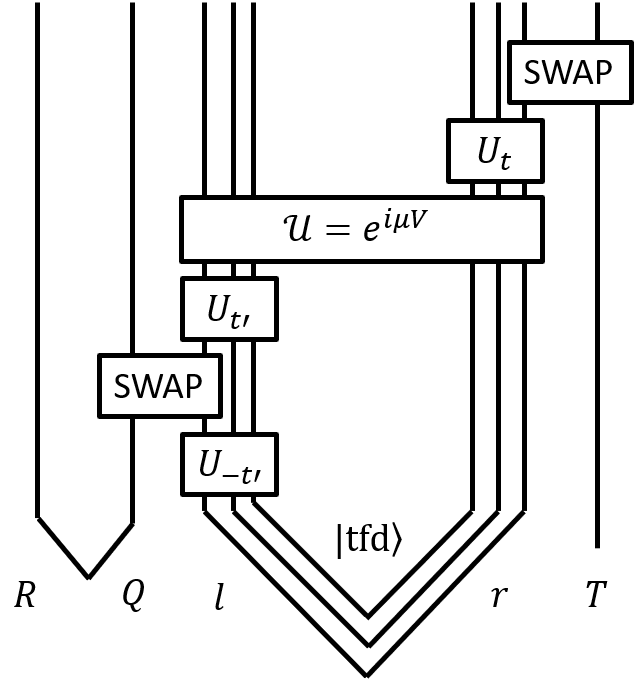}
\caption{The protocol via SWAP operators. $U_t=e^{-iHt}$ is the time evolution of SYK system. \label{fig:protocol}}
\end{figure}

Consider 5 systems: R, Q, $l$, $r$, T. Let R, Q, T be single qubit systems, where R will play the role of a 
reference,  and information
will transferred from Q to T. Let $l$ and $r$ be two identical
systems of $N/2$  qubits, where we will take $N$ to be large. 

We will consider $l$ and $r$ to be physically realized as systems of $N$ Majorana fermions, pairs of which can be identified as Dirac fermions. Taking $l$ and $r$ together there are a total of $N$ qubits. Time evolution will be generated by the SYK model Hamiltonian acting on these fermionic systems, as discussed further in the next section.

A Majorana fermion presentation of a qubit system can be constructed as follows \footnote{It would be interesting to study if these operators are easy to manipulate in some experimental realization of the SYK model.}
\begin{align}
\psi_l^{2j-1} = \f {1} {\sqrt{2}} (ZX)^{j-1}XX(II)^{N/2-j},~\psi_l^{2j} = \f {1} {\sqrt{2}} (ZX)^{j-1}YX(II)^{N/2-j} \\
\psi_r^{2j-1} = \f {1} {\sqrt{2}} (ZX)^{j-1}IY(II)^{N/2-j},~\psi_r^{2j} = \f {1} {\sqrt{2}} (ZX)^{j-1}IZ(II)^{N/2-j}
\end{align}
where $X,Y,Z,I$ stand for the Pauli matrices and 2 by 2 identity, respectively. In the above notation, each Majorana fermion is tensor product of $N$ 2 by 2 matrices acting on $N$ distinct qubits in which odd numbered qubits are assigned to the $l$ system and even numbered qubits are assigned to the $r$ system. For notational simplicity, we omit all tensor product symbols and powers $A^n$ should be interpreted in the tensor product sense $A\otimes\cdots\otimes A$. 

Initially, RQ are in a maximally entangled state
\begin{equation}
\bra{\phi_{0}}=\f 1{\sqrt{2}}(\bra{00}+\bra{11})
\end{equation}
$l$ and $r$ are in the thermofield double state
\begin{equation}
\bra{\text{tfd}}=\f 1Z\sum_{n}e^{-\b E_{n}/2}\bra{nn}_{lr}
\end{equation}
and T is in the state $\bra 0$. Success of teleportation from Q to T using the entanglement resource of $l$ and $r$ is equivalent to producing maximal entanglement
between R and T. \footnote{Maximal entanglement between R and T means that they are in one of four Bell states, which are related to each other by a Pauli matrix operation.}

The traversable wormhole protocol is as follows (see see Fig. \ref{fig:protocol}). 
\begin{enumerate}
\item At time $-t'$, apply a SWAP between Q and $l$ such that the qubit is inserted into
the wormhole.
\item At time zero, apply an interaction $\mU=e^{i\mu V}$ with $V=\mO_{l}\mO_{r}$
to the state.
\item At time $t$, apply a SWAP between T and $r$ such that the qubit is extracted 
from the wormhole.
\end{enumerate}
In the first step, we need to find an appropriate SWAP operator. Since $l$
and $r$ contain Dirac fermions, we can pick one $\chi_{l}$ and its
conjugate $\chi_{l}^{\dagger}$ with 
\begin{equation}
\{\chi_{l},\chi_{l}^{\dagger}\}=1,\;\chi_{l}^{2}=\chi_{l}^{\dagger2}=0
\end{equation}
and construct a SWAP operator as 
\begin{equation}
S_{Ql}=\begin{pmatrix}\chi_{l}\chi_{l}^{\dagger} & \chi_{l}^{\dagger}\\
\chi_{l} & \chi_{l}^{\dagger}\chi_{l}
\end{pmatrix}\label{eq:4}
\end{equation}
Note that this is the ordinary SWAP
operator, as can be seen from the  Pauli matrix
representation for $\chi_{l}$ and $\chi_{l}^{\dagger}$:
\begin{equation}
\chi_{l}\simeq\begin{pmatrix}0 & 1\\
0 & 0
\end{pmatrix},\;\chi_{l}^{\dagger}\simeq\begin{pmatrix}0 & 0\\
1 & 0
\end{pmatrix}
\end{equation}
In the final step of the protocol, we will use the
same definition for the SWAP operator $S_{Tr}$ by replacing $l$
with $r$ in (\ref{eq:4}). 

The second step can equivalently be replaced by measurement on the left together with action of a unitarity operator on the right, so that only a classical channel communication is required. The basic idea is that an interaction $e^{i \mu {\cal O}_l {\cal O}_r} = \sum_a P_a^l e^{i a \mu {\cal O}_r}$, where $P^l_a$ is the projector on to the eigensubspace of ${\cal O}_l$ with eigenvalue $a$. Whether the value of $a$ is actually measured, that is, classically recorded, on the left cannot affect the result on the right \cite{maldacena2017diving}. 

Our protocol will involve a slight generalization, in which many operators are coupled, $e^{i \mu_0 \sum_j {\cal O}_l^j {\cal O}_r^j}$, which is equivalent to a classical channel only when the  various ${\cal O}_l^j {\cal O}_r^j$ commute with each other. In particular, the construction will involve coupling the Majoranas, which can be written in terms of operations on the qubits using the above representation as
\begin{align}
i\psi_l^{2j-1}\psi_r^{2j-1}=-\f 1 2 (XZ)_j,~i\psi_l^{2j}\psi_r^{2j}=\f 1 2 (YY)_j
\end{align}
where $[\cdots]_j\equiv (II)^{j-1}[\cdots](II)^{N/2-j}$ means acting only on $j$-th left and right qubits.

It is straightforward to see that
\begin{align}
e^{i\mu_0\cdot i\psi_l^{2j-1}\psi_r^{2j-1}}&=\left[P_{x+} e^{-i\mu_0 Z/2}+P_{x-} e^{i\mu_0 Z/2}\right]_j \label{eq:P1}\\
e^{i\mu_0\cdot i\psi_l^{2j}\psi_r^{2j}}&=\left[P_{y+} e^{i\mu_0 Y/2}+P_{y-} e^{-i\mu_0 Y/2}\right]_j \label{eq:P2}
\end{align}
where $P_{x\pm}=(I\pm X)/2$ and $P_{y\pm}=(I\pm Y)/2$ are projectors onto eigenstates of $X$ and $Y$ respectively. Therefore each term in $e^{i \mu_0 \sum_j {\cal O}_l^j {\cal O}_r^j}$ is of the form of $\sum_a P_l^a U_r^a$, where each measurement is applied only on a single qubit. In standard teleportation, one usually implements a Bell pair measurement rather than a single qubit measurement  in the computational basis. The extra simplicity in our situation is possible because of the scrambling generated by time evolution  \cite{susskind2018teleportation}.

If we include Majorana fields with both $2j-1$ and $2j$ indices in the coupling $V$, their projections will not commute, as shown in \eqref{eq:P1} and \eqref{eq:P2}. Therefore, to relate to teleportation with a classical channel, one should consider coupling at most half of the $N$ Majoranas, as discussed in Sec. \ref{sec:partial}.

After the three steps, the state becomes
\begin{align}
 & S_{Tr}\mU S_{Ql}\bra{\phi_{0}}_{RQ}\bra{\text{tfd}}_{lr}\bra 0_{T}\nonumber \\
= & \f 1{\sqrt{2}}\left(\bra 0_{T}\otimes\chi_{r}\chi_{r}^{\dagger}+\bra 1_{T}\otimes\chi_{r}\right)\mU \left(\bra{00}_{RQ}\otimes\chi_{l}\chi_{l}^{\dagger}\right.\nn\\
&\left. +\bra{01}_{RQ}\otimes\chi_{l}+\bra{10}_{RQ}\otimes\chi_{l}^{\dagger}+\bra{11}_{RQ}\otimes\chi_{l}^{\dagger}\chi_{l}\right)\bra{\text{tfd}}_{lr}
\end{align}
where $\chi_r,\chi_r^\dagger$ are short for $\chi_r(t),\chi_r^\dagger(t)$ and $\chi_l,\chi_l^\dagger$ are short for $\chi_l(-t'),\chi_l^\dagger(-t')$. In the main part of this paper, we suppress their time arguments for notational simplicity.

In order to quantify the success of teleportation, we  compute the
mutual information between R and T. Perfect teleportation corresponds to
\begin{equation}
I_{RT}=S(R)+S(T)-S(RT)=2\log2\label{eq:7}
\end{equation}
The reduced density matrix of TR is
\begin{equation}
\r_{TR}=\f 12\begin{pmatrix}\r_{11} & 0 & 0 & \r_{14}\\
0 & \r_{22} & \r_{23} & 0\\
0 & \r_{23}^{*} & \r_{33} & 0\\
\r_{14}^{*} & 0 & 0 & \r_{44}
\end{pmatrix}\label{eq:17}
\end{equation}
where the trivial matrix elements are due to  fermion
number conservation 
\begin{equation}
\avg{\chi_{i_{1}}\cdots\chi_{i_{s}}\chi_{i_{s+1}}^{\dagger}\cdots\chi_{i_{2k+1}}^{\dagger}}=0,\;k\in\N
\end{equation}
and the nontrivial matrix elements are given as
\begin{align}
\r_{11} & =\avg{\chi_{l}\chi_{l}^{\dagger}\mU^\dagger\chi_{r}\chi_{r}^{\dagger}\mU\chi_{l}\chi_{l}^{\dagger}}+\avg{\chi_{l}^{\dagger}\mU^\dagger\chi_{r}\chi_{r}^{\dagger}\mU\chi_{l}}\label{eq:10}\\
\r_{14} & =\avg{\chi_{l}\mU^\dagger\chi_{r}^{\dagger}\mU\chi_{l}\chi_{l}^{\dagger}}+\avg{\chi_{l}^{\dagger}\chi_{l}\mU^\dagger\chi_{r}^{\dagger}\mU\chi_{l}}\\
\r_{22} & =\avg{\chi_{l}\mU^\dagger\chi_{r}\chi_{r}^{\dagger}\mU\chi_{l}^{\dagger}}+\avg{\chi_{l}^{\dagger}\chi_{l}\mU^\dagger\chi_{r}\chi_{r}^{\dagger}\mU\chi_{l}^{\dagger}\chi_{l}}\\
\r_{23} & =\avg{\chi_{l}\chi_{l}^{\dagger}\mU^\dagger\chi_{r}^{\dagger}\mU\chi_{l}^{\dagger}}+\avg{\chi_{l}^{\dagger}\mU^\dagger\chi_{r}^{\dagger}\mU\chi_{l}^{\dagger}\chi_{l}}\\
\r_{33} & =\avg{\chi_{l}\chi_{l}^{\dagger}\mU^\dagger\chi_{r}^{\dagger}\chi_{r}\mU\chi_{l}\chi_{l}^{\dagger}}+\avg{\chi_{l}^{\dagger}\mU^\dagger\chi_{r}^{\dagger}\chi_{r}\mU\chi_{l}}\label{eq:14}\\
\r_{44} & =\avg{\chi_{l}\mU^\dagger\chi_{r}^{\dagger}\chi_{r}\mU\chi_{l}^{\dagger}}+\avg{\chi_{l}^{\dagger}\chi_{l}\mU^\dagger\chi_{r}^{\dagger}\chi_{r}\mU\chi_{l}^{\dagger}\chi_{l}}\label{eq:15}
\end{align}
where $\mU\equiv\exp(i\mu V)$ and all expectation values are taken in thermofield double state.
A consistency check is that if $\mu=0$, then we get $\r_{TR}=\f 12\text{diag}(\r_{11},\r_{11},1-\r_{11},1-\r_{11})$
which leads to zero mutual information $I_{RT}$. 

Perfect teleportation (\ref{eq:7}) requires that both $S(R)=S(T)=2\log2$ and
$S(RT)=0$. Given the structure of $\r_{TR}$ in (\ref{eq:17}), the
first conditions imply that
\begin{equation}
\r_{11}=\r_{44},\;\r_{11}+\r_{22}=\r_{11}+\r_{33}=1\label{eq:cond1}
\end{equation}
and second condition implies that either
\begin{equation}
\begin{cases}
\r_{11}=|\r_{14}|=1,\;|\r_{23}|=0\\
\r_{11}=|\r_{14}|=0,\;|\r_{23}|=1
\end{cases}\label{eq:cond2}
\end{equation}

Consider the first case. It is obvious that 
\begin{align}
\r_{11}+\r_{33}&=\avg{\chi_{l}\chi_{l}^{\dagger}\mU^\dagger\{\chi_{r},\chi_{r}^{\dagger}\}\mU\chi_{l}\chi_{l}^{\dagger}}+\avg{\chi_{l}^{\dagger}\mU^\dagger\{\chi_{r},\chi_{r}^{\dagger}\}\mU\chi_{l}}=\avg{\{\chi_{l},\chi_{l}^{\dagger}\}}=1
\end{align}
and similarly $\r_{22}+\r_{44}=1$. Therefore,
$\r_{11}=1$ is equivalent to $\r_{33}=0$, which implies that the two
terms in (\ref{eq:14}) vanish, as both are nonnegative
\begin{equation}
\chi_{r}\mU\chi_{l}\chi_{l}^{\dagger}\bra{\text{tfd}}=\chi_{r}\mU\chi_{l}\bra{\text{tfd}}=0\label{eq:19}
\end{equation}
Similarly, $\r_{22}=0$ implies that
\begin{equation}
\chi_{r}^{\dagger}\mU\chi_{l}^{\dagger}\bra{\text{tfd}}=\chi_{r}^{\dagger}\mU\chi_{l}^{\dagger}\chi_{l}\bra{\text{tfd}}=0\label{eq:20}
\end{equation}
Based on these, $\r_{23}$ is automatically trivial. The only remaining
nontrivial condition is $|\r_{14}|=1$. From (\ref{eq:19}) and (\ref{eq:20}),
we can derive 
\begin{align}
\chi_{r}\mU\chi_{l}^{\dagger}\chi_{l}\bra{\text{tfd}}&=\chi_{r}\mU\bra{\text{tfd}},\\
\chi_{r}^{\dagger}\mU\chi_{l}\chi_{l}^{\dagger}\bra{\text{tfd}}&=\chi_{r}^{\dagger}\mU\bra{\text{tfd}}
\end{align}
which can be used to simplify  $|\r_{14}|=1$ to the condition
\begin{equation}
\left|\avg{\{\mU^\dagger\chi_{r}^{\dagger}\mU,\chi_{l}\}}\right|=1\label{eq:22}
\end{equation}
This has a clear physical interpretation: the left-right causal propagator
reaches its maximum value. Note that because we are discussing fermions, the propagator can never be larger than 1. 

For the second possible case above, the condition for perfect teleportation is similar
\begin{align}
\r_{11} & =0\implies\chi_{r}^{\dagger}\mU\chi_{l}\chi_{l}^{\dagger}\bra{\text{tfd}}=\chi_{r}^{\dagger}\mU\chi_{l}\bra{\text{tfd}}=0\\
\r_{44} & =0\implies\chi_{r}\mU\chi_{l}^{\dagger}\bra{\text{tfd}}=\chi_{r}\mU\chi_{l}^{\dagger}\chi_{l}\bra{\text{tfd}}=0\\
|\r_{23}| & =1\implies\left|\avg{\{\mU^\dagger\chi_{r}\mU,\chi_{l}\}}\right|=1
\end{align}
Comparing the two, one can see that the difference is just an exchange
of $\chi_{r}$ and $\chi_{r}^{\dagger}$. This just corresponds to flipping the convention of which state on the right we call 0 versus 1. 
%This difference basically
%depends on how do we choose the dual relation between $l$ and $r$
%systems. For example, 
If we interpret $\chi^{\dagger}$ as creation
of a particle and $\chi$ as annihilation of a particle, %we would like to take first case for $\bra{TFD}$ being in lack of a particle. From now on, we will only 
it is natural to take the first case. 

\subsection{SYK model} \label{sec:syk}
In the SYK model, we can construct Dirac fermions from two Majorana
fermions. More precisely, we will consider an even number $N$ of 
Majorana fermions on each side:
\begin{align}
H_{l,r}=i^{q/2}\sum_{1\leq j_{1}<\cdots<j_{q}\leq N}J_{j_{1}\cdots j_{q}}^{l,r}\psi_{l,r}^{j_{1}}\cdots\psi_{l,r}^{j_{q}},\\
\avg{\left(J_{j_{1}\cdots j_{q}}^{l,r}\right)^{2}}=\f{2^{q-1}\mJ^{2}(q-1)!}{qN^{q-1}}=\f{J^{2}(q-1)!}{N^{q-1}}
\end{align}
where the normalization is 
\begin{equation}\label{eq:alg}
\{\psi_{l,r}^{j},\psi_{l,r}^{k}\}=\d^{jk}
\end{equation}
Let $\bra I$ be the maximally entangled state of $l$ and $r$ defined by \begin{equation}
(\psi_{l}^{i}+i\psi_{r}^{i})\bra I=0,\quad\forall i\label{eq:34}
\end{equation}
In other words, $\bra I$ is annihilated by a complex fermion $f_{j}=\f 1{\sqrt{2}}(\psi_{l}^{j}+i\psi_{r}^{j})$.

The thermofield double state is then given
by
\begin{equation}
\bra{\text{tfd}}=Z_{\b}^{-1/2}e^{-\b H/4}\bra I,\quad H\equiv H_{l}+H_{r}
\end{equation}
We further require that $H_{l}-H_{r}\bra I=0$ so this state is the infinite temperature entangled state.
Using $(H_{l}-H_{r})\bra I=0$, this implies that the Hamiltonia must be related by
\begin{equation}
J_{j_{1}\cdots j_{q}}^{l}=i^{q}J_{j_{1}\cdots j_{q}}^{r}\label{eq:35}
\end{equation}

The SYK model can be solved using path integrals after averaging over the Gaussian ensemble of random coupling $J$. Note that the thermofield
double state is defined by a Euclidean evolution of length $\b/4$ from $\bra I$.
 Euclidean time ordered correlation functions in the thermofield
double state are related to
\begin{equation}
\avg{I|e^{-\b H/2}\T\mO_{1}\cdots\mO_{k}|I}
\end{equation}
which is equivalent to the path integral over a length of $\b/2$.\footnote{If we take all $\mO_{i}$ at $\b/4$, this is the expectation value
in the thermofield double state. } Therefore, the action we start with is 
\begin{equation}
S=\int_{0}^{\b/2}d\tau\left[\f 12\sum_{i=1}^{N}(\psi_{l}^{i}\del_{\tau}\psi_{l}^{i}+\psi_{r}^{i}\del_{\tau}\psi_{r}^{i})+H_{l}+H_{r}\right]
\end{equation}
Since we have two copies with random couplings related by
(\ref{eq:35}), integrating over $J$ will give interaction terms between
$\psi_{l}$ and $\psi_{r}$ \cite{maldacena2018eternal}.  Introducing
four delta functions
\begin{equation}
\d\left(G_{ab}(\tau_{1},\tau_{2})-\f 1N\sum_{i=1}^{N}\psi_{a}^{i}(\tau_{1})\psi_{b}^{i}(\tau_{2})\right)
\end{equation}
for $a,b=l,r$ in the path integral, we find
\begin{align}
S= & \f 12\int_{0}^{\b/2}d\tau\sum_{i=1}^{N}(\psi_{l}^{i}\del_{\tau}\psi_{l}^{i}+\psi_{r}^{i}\del_{\tau}\psi_{r}^{i})-\f 12\sum_{a,b}\int_{0}^{\b/2}d\tau d\tau'\S_{ab}(\tau,\tau')\sum_{i=1}^{N}\psi_{a}^{i}(\tau)\psi_{b}^{i}(\tau')\nonumber \\
 & +\f N2\sum_{a,b}\int_{0}^{\b/2}d\tau d\tau'\left[\S_{ab}(\tau,\tau')G_{ab}(\tau,\tau')-\f{J^{2}}qs_{ab}G_{ab}(\tau,\tau')^{q}\right]
\end{align}
where $s_{ll}=s_{rr}=1$ and $s_{lr}=s_{rl}=(-)^{q/2}$. We stitch together the left and right fermions into a single field, defining $\text{\ensuremath{\psi}}$ for $\tau\in[0,\b]$ as
\begin{equation}\label{eq:stitch}
\psi^{j}(\tau)=\begin{cases}
\psi_{l}^{j}(\tau) & \tau\in[0,\b/2]\\
i\psi_{r}^{j}(\b-\tau) & \tau\in[\b/2,\b]
\end{cases}
\end{equation}
and corresponding $G$, $\S$ fields
\begin{align}
&G(\tau,\tau'),\S(\tau,\tau')\nn\\
=&\begin{cases}
G_{ll}(\tau,\tau'),\S_{ll}(\tau,\tau') & \tau,\tau'\in[0,\b/2]\\
iG_{lr}(\tau,\tau'_\b),-i\S_{lr}(\tau,\tau'_\b) & \tau,\tau'_\b\in[0,\b/2]\\
iG_{rl}(\tau_\b,\tau'),-i\S_{rl}(\tau_\b,\tau') & \tau_\b,\tau'\in[0,\b/2]\\
-G_{rr}(\tau_\b,\tau'_\b),-\S_{rr}(\tau_\b,\tau'_\b) & \tau_\b,\tau'_\b\in[0,\b/2]
\end{cases}
\end{align}
where $\tau_\b\equiv \b-\tau$. The action then simplifies to 
\begin{align}
S=&\f 12\sum_{i=1}^{N}\left[\int_{0}^{\b}d\tau\psi^{i}\del_{\tau}\psi^{i}-\int_{0}^{\b}d\tau d\tau'\psi^{i}(\tau)\S(\tau,\tau')\psi^{i}(\tau')\right]\nn\\
&+\f N2\int_{0}^{\b}d\tau d\tau'\left[\S(\tau,\tau')G(\tau,\tau')-\f{J^{2}}qG(\tau,\tau')^{q}\right]
\end{align}
which unsurprisingly becomes the path integral of a single field $\psi$ on the thermal circle of
length $\b$. The only thing to check in this treatment is that on the
stitching boundary $\b/2$, the field redefinition may not be smooth.
Fortunately, if we extend the definition of $\psi_{l,r}^{j}$ beyond
$\tau=\b/2$ via Euclidean evolution, we have 
\begin{align}
&\ket Ie^{-\b H/2}\psi_{l}^{j}(\b/2+\tau)=\ket Ie^{H\tau}\psi_{l}^{j}e^{-(\b/2+\tau)H}\nn\\
=&\ket I\psi_{l}^{j}e^{2H_{r}\tau}e^{-(\b/2+\tau)H}=i\ket Ie^{-\b H/2}\psi_{r}^{j}(\b/2-\tau)
\end{align}
thanks to (\ref{eq:34}). This shows that the fields in the path integral are smoothly
defined at $\tau=\b/2$. Moreover, one can similarly show that 
\begin{equation}
\psi_{l}^{j}(-\tau)\bra I=-i\psi_{r}^{j}(\tau)\bra I
\end{equation}
which means that we can even extend the path integral past $\tau=\b$
by requiring the usual thermal anti-periodic $\psi^{i}(\tau)=-\psi^{i}(\b+\tau)$ boundary condition.
Integration over $\psi^{i}$ results in an effective action for the $G$ and
$\S$ fields.
\begin{equation}
S[G,\S]=-\f N2\log\det(\del_{\tau}-\S)+\f N2\iint(\S G-\f 1qJ^{2}G^{q})\label{eq:eff-act}
\end{equation}
In the large $N$ limit, the equation of motion is
\begin{align}\label{eq:91}
G^{-1}=G_{0}^{-1}-&\S,\;\S(\tau,\tau')=J^{2}G(\tau,\tau'){}^{q-1},\\
G_{0}^{-1}=\del_{\tau}&\iff G_{0}(\tau)=\f 12\sgn\tau
\end{align}

%Such stitching process can be interpreted in another perspective. Algebraically we can always define stitching fields above after we choose a specific relation between the random couplings of two systems like (\ref{eq:35}). After smearing it, the path integral seems to know that the best optimized state is $\bra I$ because the leading order contribution should always come from a smooth solution. 

We will add instantaneous interactions between the two SYK systems at time $\tau_{0}$
by \cite{maldacena2018eternal}
\begin{equation}
H_{int}=-\mu V=-\mu\times\f i{qN}\sum_{j=1}^{N}\psi_{l}^{j}\psi_{r}^{j}=-\f{\mu}{qN}\sum_{j=1}^{N}f_{j}^{\dagger}f_{j}+\f{\mu}{2q}\label{eq:36}
\end{equation}
We will take $\mu$ to be $O(1)$ in the $1/N$ counting, as in previous work \cite{maldacena2017diving,gao2018regenesis}. Thus $\mu V$ has an $O(1)$
effect on the state $\bra I$, namely
\begin{equation}
\mu V\bra I=-\f{\mu}{2q}\bra I
\end{equation}
 Writing $H_{int}$ in terms of $G$ field, the action is shifted by
\begin{equation}\label{eq:ds}
\d S=-\f{\mu}{2q}\left[G(\tau_{0},\b-\tau_{0})-G(\b-\tau_{0},\tau_{0})\right]
\end{equation}
%As this term is not proportional to $N$, 
Such an instantaneous interaction 
does not affect the equation of motion but only changes the boundary
condition of $G$ and $\S$ at $\tau_{0}$ (see Sec. \ref{subsec:Twist-boundary-condition}).

% (Note: I don't think there is really a difference of this type. Instantaneous interactions can only change boundary terms because they appear at that time only. If the interaction is order 1, it just makes a order 1/N change in the boundary condition. Or are you saying it's different here because the equations of motion are non-local? What I'm saying is, why don't you call 2.45 also just a change in boundary conditions, because of the delta functions?) \PG{I agree. Let us use the same language of saying it only changes boundary condition. And this will be more relating to the complete version of twist boundary condition.}

%However, if $\mu$ were chosen as order $N$, and it would enter equation
%of motion as
It turns out that
\begin{align}
\S(\tau,\tau')=&J^{2}G(\tau,\tau'){}^{q-1}+\f{\mu}{qN}\left[\d_{\tau,\tau_{0}}\d_{\tau',\b-\tau_{0}}-\d_{\tau,\b-\tau_{0}}\d_{\tau',\tau_{0}}\right]
\end{align}
where $\d_{a,b}$ is short for $\d(a-b)$. On the other hand, $G^{-1}=G_{0}^{-1}-\S$ can be written as
\begin{equation}
\del_{\tau'}G(\tau',\tau)-\int d\tau''\S(\tau',\tau'')G(\tau'',\tau)=\d(\tau-\tau')
\end{equation}
Integrating $\tau'$ around $\tau_{0}$ and $\b-\tau_{0}$ respectively,
we get
\begin{align}
G(\tau_{0+},\tau)-&G(\tau_{0-},\tau) =\f{\mu}{qN}G(\b-\tau_{0},\tau)\label{eq:2.47}\\
G((\b-\tau_{0})_{-},\tau)-&G((\b-\tau_{0})_{+},\tau) =\f{\mu}{qN}G(\tau_{0},\tau)\label{eq:2.48}
\end{align}
where subscript in $\tau_\pm$ means $\tau\pm\varepsilon$ for $\varepsilon\ra 0$. Note that the RHS of above equations are not defined {\it a priori}, since $\tau_0$ and $\b-\tau_0$ are singular points. Instead these two equations should be taken as the definition of the values of $G$ at these two points, as follows. Consider the expansion of $G$  as a series in powers of $\mu$, so at leading order % and when $\mu=0$ we should expect continuity of 
$G$ is continuous. Hence, the discontinuity of $G$ around $\tau_0$ and $\b-\tau_0$ at linear order in $\mu$ is given by the zero-th order value of $G$ on the RHS.\footnote{Beyond linear order, it is not clear how to determine the discontinuity directly from \eqref{eq:2.47} and \eqref{eq:2.48}.} This is exactly the linearized version of twist boundary condition of $G$ we will derive from the microscopic fermion description 
in Sec. \ref{subsec:Twist-boundary-condition}.

\subsection{Density matrix $\protect\r_{TR}$ in SYK model} \label{sec:2.3}

For each of the pair of SYK systems, we construct Dirac fermions by combining two
Majorana fermions
\begin{equation}
\chi_{l,r}^{j}=\f 1{\sqrt{2}}(\psi_{l,r}^{2j-1}+i\psi_{l,r}^{2j}),\;\chi_{l,r}^{j\dagger}=\f 1{\sqrt{2}}(\psi_{l,r}^{2j-1}-i\psi_{l,r}^{2j})
\end{equation}
with commutation relations
\begin{equation}
\{\chi_{l,r}^{i},\chi_{l,r}^{j\dagger}\}=\d^{ij}
\end{equation}
It follows that
\begin{equation}
\chi_{l,r}^{j}\chi_{l,r}^{j\dagger}=\f 12-i\psi_{l,r}^{2j-1}\psi_{l,r}^{2j},~\chi_{l,r}^{j\dagger}\chi_{l,r}^{j}=\f 12+i\psi_{l,r}^{2j-1}\psi_{l,r}^{2j}
\end{equation}

Using above constructions, we can write down the matrix element of
$\r_{TR}$ more explicitly. In this subsection, we take $\chi_{l,r}^{1}$ as the Dirac fermion appearing in the general protocol described in section \ref{sec:setup}. Take $\r_{11}$ as an example. The first term
is
\begin{align}
\avg{\chi_{l}^{1}\chi_{l}^{1\dagger}\mU^\dagger\chi_{r}^{1}\chi_{r}^{1\dagger}\mU\chi_{l}^{1}\chi_{l}^{1\dagger}}= & \f 12\avg{\chi_{l}^{1}\chi_{l}^{1\dagger}}-\f i4\avg{\mU^\dagger\psi_{r}^{1}\psi_{r}^{2}\mU}-\f 12\avg{\{\psi_{l}^{1}\psi_{l}^{2},\mU^\dagger\psi_{r}^{1}\psi_{r}^{2}\mU\}}\nn\\
&+i\avg{\psi_{l}^{1}\psi_{l}^{2}\mU^\dagger\psi_{r}^{1}\psi_{r}^{2}\mU\psi_{l}^{1}\psi_{l}^{2}}\label{eq:40}
\end{align}
The second term is
\begin{align}
\avg{\chi_{l}^{1\dagger}\mU^\dagger\chi_{r}^{1}\chi_{r}^{1\dagger}\mU\chi_{l}^{1}}= & \f 12\avg{\chi_{l}^{1\dagger}\chi_{l}^{1}}-\f i2\avg{\psi_{l}^{1}\mU^\dagger\psi_{r}^{1}\psi_{r}^{2}\mU\psi_{l}^{1}}-\f 12\avg{\psi_{l}^{2}\mU^\dagger\psi_{r}^{1}\psi_{r}^{2}\mU\psi_{l}^{1}}\nonumber \\
 & +\f 12\avg{\psi_{l}^{1}\mU^\dagger\psi_{r}^{1}\psi_{r}^{2}\mU\psi_{l}^{2}}-\f i2\avg{\psi_{l}^{2}\mU^\dagger\psi_{r}^{1}\psi_{r}^{2}\mU\psi_{l}^{2}}\label{eq:41}
\end{align}

For the SYK model, after performing the ensemble average, there is an emergent
$SO(N)$ symmetry that rotates all $\psi^{i}_{l,r}$ by $U^{ij}\psi^{j}_{l,r}$.
This symmetry must be manifested by correlation functions, thus
\begin{equation}
\avg{\psi^{i_{1}}\cdots\psi^{i_{2k}}}\propto\d^{\pi_{1}\pi_{2}}\cdots\d^{\pi_{2k-1}\pi_{2k}}
\end{equation}
where $\pi_{i}$ is the permutation of $\{i_{1},\cdots,i_{2k}\}$.
As $V$ contains all pairs of $\psi_{l}^{i}$ and $\psi_{r}^{i}$
with the same indices, the correlation functions in (\ref{eq:40}) and
(\ref{eq:41}) with an odd number of $1$ and $2$ indices vanish. Moreover,
this symmetry dictates that correlation functions are invariant
under exchange of the $1$ and $2$ indices. Therefore, we get 
\begin{equation}
\r_{11}=\f 12\left[1-\avg{\{\psi_{l}^{1},[\psi_{l}^{2},\mU^\dagger\psi_{r}^{1}\psi_{r}^{2}\mU]\}}\right]
\end{equation}

Similarly, we can work out all other matrix elements of $\r_{TR}$.
They are 
\begin{align}
\r_{14} & =\f 12\avg{\{\psi_{l}^{1},\mU^\dagger\psi_{r}^{1}\mU\}}-\avg{\{\psi_{l}^{2},\psi_{l}^{1}\mU^\dagger\psi_{r}^{2}\mU\psi_{l}^{1}\}}\label{eq:46-0}\\
\r_{22} & =\f 12\left[1+\avg{\{\psi_{l}^{1},[\psi_{l}^{2},\mU^\dagger\psi_{r}^{1}\psi_{r}^{2}\mU]\}}\right]=1-\r_{11}\\
\r_{23} & =0,\quad\r_{33}=1-\r_{11},\quad\r_{44}=1-\r_{22}\label{eq:46}
\end{align}
Note that in (\ref{eq:46}) $\r_{23}$ becomes trivial exactly due
to the $SO(N)$ symmetry, while $\r_{33}$ and $\r_{44}$ are related to $\r_{11}$
and $\r_{22}$ by definition. Comparing with the conditions for perfect teleportation (\ref{eq:cond1}) and
(\ref{eq:cond2}), the only nontrivial conditions  for obtaining maximal mutual
information between R and T are
\begin{equation}
\r_{11}=|\r_{14}|=1\label{eq:47}
\end{equation}

\section{Computing the correlation functions in the SYK model}\label{sec:solve}

In this section, we will use the method developed in \cite{qi2018quantum}.
To calculate $\r_{11}$ and $\r_{14}$, there are two types of correlation
functions that need to be found
\begin{equation}
\avg{e^{-i\mu V}\psi^{1}e^{i\mu V}\psi^{1}},\;\avg{e^{-i\mu V}\psi^{1}\psi^{2}e^{i\mu V}\psi^{1}\psi^{2}}
\end{equation}
where we will consider all orderings of the operators.
%do not specify the ordering of each operator but simply put all orderings into these two types. 
We first focus on the former. For notational simplicity, in this section we use $\psi$ to represent $\psi^1$.

\subsection{Stitching correlation function}

With the stitching field $\psi$ defined in \eqref{eq:stitch}, we define a stitching correlation
function for $\tau_{a},\tau_{b}\in[0,\b]$ as
\begin{equation}
\mG_{\mu}(\tau_{a},\tau_{b})=\f{\avg{I|e^{-\b H/2}\bar{\mT}\left[e^{-\mu V(\b/4+\e)}e^{\mu V(\b/4-\e)}\psi(\tau_{a})\psi(\tau_{b})\right]|I}}{\avg{I|e^{-\b H/2}e^{-\mu V(\b/4+\e)}e^{\mu V(\b/4-\e)}|I}}\sgn(\bar{\tau}_{a}-\bar{\tau}_{b})\label{eq:3}
\end{equation}
where $\bar{\tau}\equiv\min(\tau,\b-\tau)$ and $\bar{\mT}$ represents time ordering for $\bar{\tau}_{a}$ and $\bar{\tau}_{b}$
(without any $\pm$ sign from (anti-)commutation of $\psi$). Here
$\e\in(0,\b/4)$ is a regulator that we will ultimately take to zero. As discussed in Sec. \ref{sec:syk}, this correlation function is smoothly defined around $\b/2$. Note that $\mG_\mu$ clearly can be extended to any real $\tau_{a}$ and
$\tau_{b}$ using the periodicity of $\psi$, $\psi(\tau+\b)=-\psi(\tau)$. We will call $\tau_{a},\tau_{b}\in[0,\b]$ the basic domain (see Fig. \ref{fig:The-stitching-correlation}).

\subsubsection{Symmetries\label{subsec:Symmetries}}

There are several symmetries of $\mG_\mu$. It
is anti-symmetric in $\tau_{a}$ and $\tau_{b}$
\begin{equation}
\mG_{\mu}(\tau_{a},\tau_{b})=-\mG_{\mu}(\tau_{b},\tau_{a})\label{eq:4-1}
\end{equation}
which together with the periodicity can be summarized as
\begin{align}
\mG_{\mu}(\tau_{a},\tau_{b})=-\mG_{\mu}(\tau_{b},\tau_{a})=(-)^{n+m}\mG_{\mu}(\tau_{a}+n\b,\tau_{b}+m\b)
\end{align}

Another feature is that $\mG_{\mu}$ is real. This can be seen
as follows. First, the Hamiltonia $H_{l,r}$ have only real coefficients
and $\mu$ is real. Second, $\psi_{l}^{j}=\f 1{\sqrt{2}}(f_{j}+f_{j}^{\dagger})$
and $i\psi_{r}^{j}=\f 1{\sqrt{2}}(f_{j}-f_{j}^{\dagger})$ are linear combinations with real
coefficients. Hence we can easily conclude that as operators 
both $e^{-\b H/2}\bar{\mT}\left[e^{-\mu V(\b/4+\e)}e^{\mu V(\b/4-\e)}\psi(\tau_{a})\psi(\tau_{b})\right]$
and $e^{-\b H/2}e^{-\mu V(\b/4+\e)}e^{\mu V(\b/4-\e)}$ 
consist of $c_{i}$ and $c_{i}^{\dagger}$ with real coefficients.
Thus $\mG_{\mu}$ must be real.

This leads to further symmetries. Define $\mO^{l}=\psi_l$, $\mO^r=i\psi_r$ and 
\begin{align} 
\mC^{AB}_{\pm}(\tau_1,\tau_2)=\mT\left[e^{-\mu V(\b/4\pm\e)}e^{\mu V(\b/4\mp\e)}\mO^{A}(\tau_1)\mO^{B}(\tau_2)\right]
\end{align}
For $\bar{\tau}_{a},\bar{\tau}_{b}\in[0,\b/2]$ and $A,B=l,r$, we have
\begin{align}
&\avg{I|e^{-\b H/2}\mC_+^{AB}(\bar{\tau}_{a},\bar{\tau}_{b})|I}^*=(-1)^s\avg{I|e^{-\b H/2}\mC_-^{AB}(\b/2-\bar{\tau}_{a},\b/2-\bar{\tau}_{b})|I}
\end{align}
where $s=0$ if $A=B$ and $s=1$ if $A\neq B$ (because there is
an $i$ with $\psi_{r}$), and $\mT$ is ordinary time ordering. Here
we used the fact
\begin{equation}
\mO(\tau)^{\dagger}=\left[e^{H\tau}\mO e^{-H\tau}\right]^{\dagger}=e^{-H\tau}\mO e^{H\tau}=\mO(-\tau)
\end{equation}
This implies that in the basic domain $\tau_{a},\tau_{b}\in[0,\b]$
\begin{equation}
\mG_{\mu}(\tau_{a},\tau_{b})=\mG_{-\mu}(\b/2-\tau_{b},\b/2-\tau_{a})\label{eq:22-1}
\end{equation}
Using this symmetry and  shifts by $\b$, we get another symmetry
\begin{align}
\mG_{\mu}(\tau_{a},\tau_{b}) & =\mG_{-\mu}(3\b/2-\tau_{a},\b/2-\tau_{b})\label{eq:23}\\
 & =\mG_{-\mu}(3\b/2-\tau_{b},3\b/2-\tau_{a})\label{eq:24}
\end{align}

Finally, if $q\in4\Z$ there is one more discrete symmetry of $\mG_{\mu}$ resulting from 
\begin{equation}
\psi_{l}^{j}\ra\psi_{r}^{j},\;\psi_{r}^{j}\ra-\psi_{l}^{j}\label{eq:77}
\end{equation}
This can be seen by noting that $\mG_{\mu}$ is the expectation of a complicated
operator that can be expanded as a sum of various products of $\psi_{l}^{j}$
and $\psi_{r}^{k}$ in the state $\bra I$. Since $\bra I$ is defined
by $(\psi_{l}^{j}+i\psi_{r}^{j})\bra I=0$ for all $j$, $\bra I$
has a tensor product form  $\otimes_{j}\bra{I_{j}}$. Therefore,
we only need to check if all correlations in each $\bra{I_{j}}$ are
invariant under (\ref{eq:77}). It is clear that for all $j$
\begin{align}
\avg{I_{j}|\psi_{l}^{j}\psi_{r}^{j}|I_{j}}=\f i2,~\avg{I_{j}|\psi_{l}^{j}|I_{j}}=\avg{I_{j}|\psi_{r}^{j}|I_{j}}=0
\end{align}
and all higher point correlations can be reduced to them using \eqref{eq:alg}. This means
that whatever $\mG_{\mu}$ is, the whole expression is invariant under
(\ref{eq:77}). Second, $V$ is invariant
\begin{equation}
V=i\sum_{j}\psi_{l}^{j}\psi_{r}^{j}=i\sum_{j}\psi_{r}^{j}(-\psi_{l}^{j})=i\sum_{j}\psi_{l}^{j}\psi_{r}^{j}
\end{equation}
Third the Hamiltonian $H$ is invariant when $q\in4\Z$ as then $J^{l}=J^{r}$. Hence we have the following identity
\begin{equation}
\mG_{\mu}(\tau_{a},\tau_{b})=-\mG_{\mu}(\b-\tau_{a},\b-\tau_{b})=\mG_{\mu}(\b-\tau_{b},\b-\tau_{a})\label{eq:29}
\end{equation}

\begin{figure}
\centering
\includegraphics[width=7.4cm]{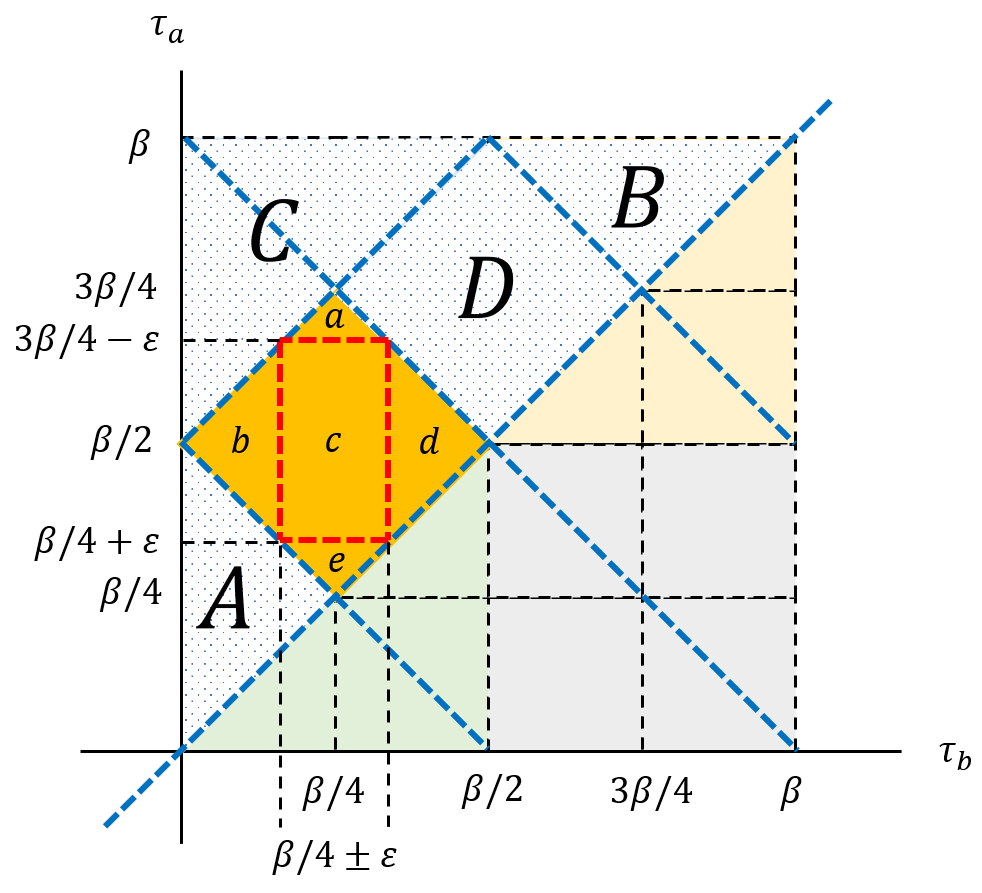}
\caption{The stitching correlation function $\mG_{\mu}$ in the basic domain is split into a few small regions. Due to symmetries, the independent region is the dark yellow one, which we call the
fundamental region. Blue dashed lines are invariant
under symmetries of $\protect\mG_{\mu}$, which implies that the white
dotted regions $A,B,C,D$ are not independent. The twisted boundary
condition is on the four red lines, which split the fundamental domain
further into five subregions $a,b,c,d,e$. \label{fig:The-stitching-correlation}}
\end{figure}

These symmetries kill some regions in the left top triangle in the
basic domain. Indeed, (\ref{eq:22-1}) and (\ref{eq:24}) says $\tau_{a}+\tau_{b}=\b/2,\;3\b/2$
are two invariant lines, which kill the regions $A$ and $B$ in Fig. \ref{fig:The-stitching-correlation}.
(\ref{eq:23}) implies the existence of a reflection symmetry around point $(\tau_{a},\tau_{b})=(3\b/4,\b/4)$
which kills region $C$. (\ref{eq:29}) says $\tau_{a}+\tau_{b}=\b$
is an invariant line, which kills region $D$. In the end, the only
independent region is the dark yellow square in Fig. \ref{fig:The-stitching-correlation}, which we will call the fundamental region. 

\subsubsection{Twist boundary condition\label{subsec:Twist-boundary-condition}}

The boundary conditions on the fermions induced by the instantaneous interaction result from the following identity
\begin{equation}\label{eq:3.14}
e^{\mu V}\begin{pmatrix}\psi_{l}\\
i\psi_{r}
\end{pmatrix}e^{-\mu V}=\begin{pmatrix}\cosh\f{\mu}{qN} & -\sinh\f{\mu}{qN}\\
-\sinh\f{\mu}{qN} & \cosh\f{\mu}{qN}
\end{pmatrix}\begin{pmatrix}\psi_{l}\\
i\psi_{r}
\end{pmatrix}
\end{equation}
When $\psi$ moves across $\b/4\pm\e$, it gets acted upon by the above connection
equation (and its $-\mu$ version). It follows that for the stitching
field $\psi$
\begin{align}
&\lim_{\tau\ra(\b/4\mp\e)_{-}}\begin{pmatrix}\psi(\tau)\\
\psi(\b-\tau)
\end{pmatrix}=\begin{pmatrix}\cosh\f{\mu}{qN} & \mp\sinh\f{\mu}{qN}\\
\mp\sinh\f{\mu}{qN} & \cosh\f{\mu}{qN}
\end{pmatrix}\lim_{\tau\ra(\b/4\mp\e)_{+}}\begin{pmatrix}\psi(\tau)\\
\psi(\b-\tau)
\end{pmatrix}
\end{align}
As $\mG_{\mu}$ contains two $\psi$ insertions, this gives a twist boundary
condition
\begin{align}
&\begin{pmatrix}\underset{\tau_{a},\tau_{b}\ra(\b/4\mp\e)_{+}}{\lim}\mG_{\mu}(\tau_{a},\tau_{b})\\
\underset{\tau_{a},\tau_{b}\ra(3\b/4\pm\e)_{-}}{\lim}\mG_{\mu}(\tau_{a},\tau_{b})
\end{pmatrix}=\begin{pmatrix}\cosh\f{\mu}{qN} & \pm\sinh\f{\mu}{qN}\\
\pm\sinh\f{\mu}{qN} & \cosh\f{\mu}{qN}
\end{pmatrix}\begin{pmatrix}\underset{\tau_{a},\tau_{b}\ra(\b/4\mp\e)_{-}}{\lim}\mG_{\mu}(\tau_{a},\tau_{b})\\
\underset{\tau_{a},\tau_{b}\ra(3\b/4\pm\e)_{+}}{\lim}\mG_{\mu}(\tau_{a},\tau_{b})
\end{pmatrix}
\end{align}
Using the symmetries we can rewrite everything in terms of twist boundary conditions in fundamental
region. Indeed, we have
\begin{align}
& \mG_{\mu}(\tau_{a},(3\b/4\pm\e)_{-})=-\mG_{\mu}(\b-\tau_{a},(\b/4\mp\e)_{+})\label{eq:85}\\
& \mG_{\mu}((\b/4-\e)_{\pm},\tau_{b}) =-\mG_{-\mu}((\b/4+\e)_{\mp},\b/2-\tau_{b})\label{eq:86}\\
& \mG_{\mu}((3\b/4+\e)_{\mp},\tau_{b}) =\mG_{-\mu}((3\b/4-\e)_{\pm},\b/2-\tau_{b})\label{eq:87}
\end{align}
It follows that the independent twist boundary conditions in fundamental region are
\begin{align}
&\begin{pmatrix}\mG_{\mu}((\b/4+\e)_{+},\tau_{b})\\
\mG_{\mu}((3\b/4-\e)_{-},\tau_{b})
\end{pmatrix}=\begin{pmatrix}\cosh\f{\mu}{qN} & -\sinh\f{\mu}{qN}\\
-\sinh\f{\mu}{qN} & \cosh\f{\mu}{qN}
\end{pmatrix}\begin{pmatrix}\mG_{\mu}((\b/4+\e)_{-},\tau_{b})\\
\mG_{\mu}((3\b/4-\e)_{+},\tau_{b})
\end{pmatrix}\label{eq:tbd-2}
\end{align}
and 
\begin{align}
\mG_{\mu}(\tau_{a},& (\b/4\mp\e)_{+})=\cosh\f{\mu}{qN}\mG_{\mu}(\tau_{a},(\b/4\mp\e)_{-})\mp\sinh\f{\mu}{qN}\mG_{\mu}(\b-\tau_{a},(\b/4\mp\e)_{-})\label{eq:tbd-1}
\end{align}
They are on the red dashed lines in Fig. \ref{fig:The-stitching-correlation}, and
separate the fundamental region into five subregions $a,b,c,d,e$. One can check that at linear order in $1/N$, the twist boundary conditions are the same as \eqref{eq:2.47} and \eqref{eq:2.48}.

\subsubsection{Smoothness and UV conditions}

\begin{figure}
\begin{centering}
\includegraphics[width=8cm]{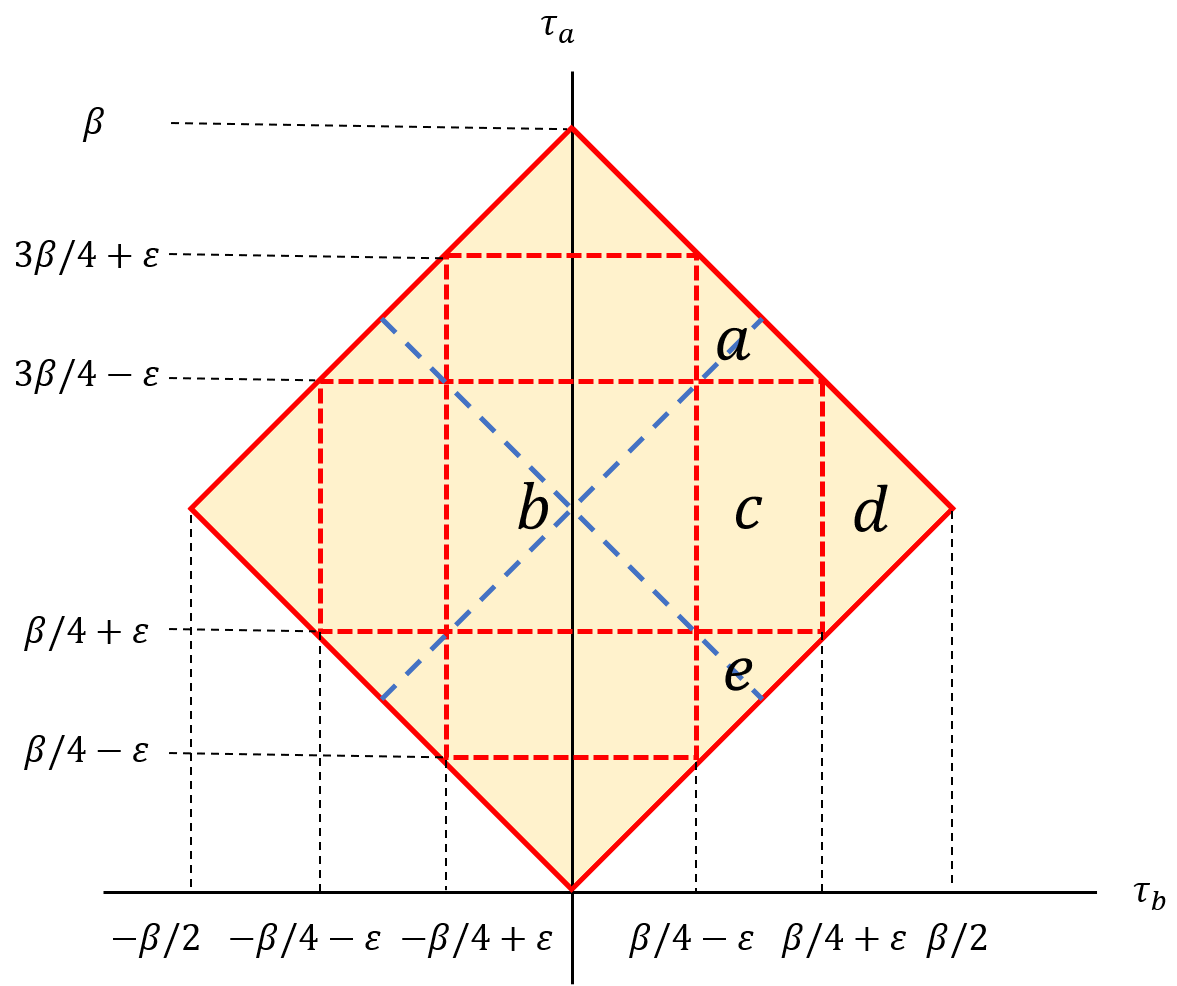}
\par\end{centering}
\caption{The smoothness condition for $\protect\mG_{\mu}$. $\protect\mG_{\mu}$
needs to be smooth in each region bounded by dashed and solid red
lines. On dashed red lines, $\protect\mG_{\mu}$ must satisfy a twist
boundary condition, and on solid red lines, $\protect\mG_{\mu}$ need
not be smooth. Blue dashed lines are symmetry lines on which
$\protect\mG_{\mu}$ must be smooth. Due to the symmetry, we only
need to solve for $\protect\mG_{\mu}$ on the five regions $a,b,c,d,e$.
\label{fig:The-smoothness-condition}}
\end{figure}
Although the symmetries reduced the problem
to the fundamental region, that does not mean that we can take an arbitrary function
that satisfies the twist boundary conditions and simply transform it to the other
regions via the symmetry. In particular, the correlation function $\mG_{\mu}$
should be further constrained by smoothness and UV conditions.

The smoothness condition is that $\mG_{\mu}$ must be a smooth function
of $\tau_{a,b}$ when no operators coincide with each other. It follows
that $\mG_{\mu}$ must be smooth crossing certain symmetry lines, for
example, the lines $\tau_{a}=\pm\tau_{b}+\b/2$. On the other hand,
some symmetry lines are ``hard'' lines that correspond to coincident operators, for example, $\tau_{a}=\pm\tau_{b}$ and $\tau_{a}=\pm\tau_{b}+\b$.
In the latter case, $\mG_{\mu}$ need not be
smooth and the extension across them is simply to  ``copy and paste''
by the symmetry.

In doing this, we need to study $\mG_{\mu}$ on a
bigger patch as shown in Fig. \ref{fig:The-smoothness-condition}.
In this figure, $\mG_{\mu}$ needs to be smooth in each region bounded
by dashed and solid red lines. On dashed red lines, $\mG_{\mu}$ must
satisfy the twist boundary condition, and on solid red lines, $\mG_{\mu}$
need not be smooth. Blue dashed lines are symmetry lines on
which $\mG_{\mu}$ must be smooth.

The input from the UV is that when $\tau_{a}\ra\tau_{b}$, $\mG_{\mu}$
must equal $1/2$, which are required by the Majorana fermion algebra. This
gives a boundary condition on the line $\tau_{a}=\tau_{b}$.

Due to the symmetry, we only need to solve for $\mG_{\mu}$ on
the new five subregions $a,b,c,d,e$ in Fig. \ref{fig:The-smoothness-condition}.

\subsection{Large $q$ solution}

\subsubsection{General solution}

The insertion of $e^{\pm \mu V}$ in the correlation function $\mG_{\mu}$
can be understood as adding a deformation to action at $\b/4\pm\e$ of the form of \eqref{eq:ds}. %As we choose $\mu$ to be order 1, it 
This does not change the bulk equation of motion \eqref{eq:91}, but simply imposes the twist boundary condition we analyzed above. 

If we fix $\mJ$, the large $q$ limit will
truncate (\ref{eq:91}) to first order \cite{eberlein2017quantum, gross2017generalization, maldacena2016remarks} as 
\begin{equation}
[G]=[G_{0}]+[G_{0}][\S][G_{0}],\quad\S=\f{\mJ^{2}}q(2G)^{q-1}\label{eq:96}
\end{equation}
where $[A][B]$ means product by integrating over the intermediate variable $\tau$. We make the ansatz for $\tau_{ab}>0$
\begin{align}
G(\tau_{a},\tau_{b})&=G_{0}(\tau_{ab})e^{\s(\tau_{a},\tau_{b})/(q-1)}\implies\S(\tau_{a},\tau_{b})=\f{\mJ^{2}}qe^{\s(\tau_{a},\tau_{b})}
\end{align}
Taking time derivative of $[G]$ in (\ref{eq:96}), we get
\begin{equation}
\del_{a}\del_{b}(G-G_{0})=-\f{\mJ^{2}}qe^{\s(\tau_{a},\tau_{b})},\;\tau_{ab}>0
\end{equation}
In the large $q$ limit, this becomes
\begin{equation}
\del_{a}\del_{b}\s=-2\mJ^{2}e^{\s(\tau_{a},\tau_{b})}\label{eq:100}
\end{equation}
which is the well-known Liouville equation. The most general solution
is
\begin{equation}
\s=\ln\f{f'(\tau_{a})g'(\tau_{b})}{(1+\mJ^{2}f(\tau_{a})g(\tau_{b}))^{2}}\label{eq:103-1}
\end{equation}
It follows that the general solution for $G$ in the large $q$ limit is
\begin{equation}
G(\tau_{a},\tau_{b})=G_{0}(\tau_{ab})\left[\f{f'(\tau_{a})g'(\tau_{b})}{(1+\mJ^{2}f(\tau_{a})g(\tau_{b}))^{2}}\right]^{\f 1{q-1}}\label{eq:103}
\end{equation}

This solution has a global $SL(2)$ symmetry. Under the following
transformation 
\begin{equation}
f(x)\ra\f{a+bf(x)}{c+df(x)},\;g(x)\ra\f{d-c\mJ^{2}g(x)}{\mJ^{2}(-b+a\mJ^{2}g(x))},\label{eq:global-sl2}
\end{equation}
where $bc-ad=1$, (\ref{eq:103-1}) is invariant.\footnote{Note that overall scaling of $a,b,c,d$ does not change $f$, and thus
we only the $SL(2)$ subgroup of $GL(2)$ acts non-trivially.} This global $SL(2)$ symmetry is a gauge redundancy in the expression
of $G$. In the following, we will discuss the solution for $G$ modulo
this redundancy.

\subsubsection{Translation invariant solution: $\mu=0$ case}

When $\mu=0$, $\mG_{\mu}=G$ becomes the usual time translation invariant thermal correlation function. This means that $G(\tau_{a},\tau_{b})$
is a function of $\tau=\tau_{ab}$ only. In particular, $\del_{b}=-\del_{a}$
in the differential equation (\ref{eq:100}) and we get
\begin{equation}
\del_{\tau}^{2}\s(\tau)=2\mJ^{2}e^{\s(\tau)}
\end{equation}
The general solution to this is
\begin{equation}
e^{\s(\tau)}=\f{\w^{2}}{\mJ^{2}\cos^{2}(\w\tau+v)}\label{eq:107}
\end{equation}
where $\w$ and $v$ are two integral constants. If we set $\tau_{b}=0$
and $\tau_{a}=\tau$, we can compare it with (\ref{eq:103-1}) and
find
\begin{equation}
f(\tau)=\f{A+B\tan(\w\tau+v)}{C+D\tan(\w\tau+v)}
\end{equation}
where $A,B,C,D$ are some constants involving $g(0)$. We fix the $SL(2)$ symmetry by choosing 
\begin{equation}
f(\tau)=\f 1{\mJ}\tan(\w\tau+v_{1})
\end{equation}
Assume $g(\tau)=\f 1{\mJ}\tan\w\tilde{g}(\tau)$. Plugging this into (\ref{eq:103-1})
 we find
\begin{equation}
e^{\s(\tau_{ab})}=\f{\w^{2}\tilde{g}'(\tau_{b})}{\mJ^{2}\cos^{2}(\w\tau_{a}-\w\tilde{g}(\tau_{b})+v_{1})}
\end{equation}
Comparing with (\ref{eq:107}), it is easy to see that $\tilde{g}(\tau_{b})=\tau_{b}+v_{2}$
for some constant $v_{2}$. 

In the following, Sec. \ref{sec:large-q}, and Appendix \ref{app}, we will shift $\tau_{a}$ to $\tau_{a}\ra\tau_{a}+\b/2$
to simplify the action of the symmetries. In these shifted coordinates,
the symmetries discussed in subsection \ref{subsec:Symmetries} now
become (as $\mu=0$)
\begin{equation}
(f(\tau_{a}),g(\tau_{b}))\simeq(g(\tau_{a}),f(\tau_{b}))\simeq(g(-\tau_{a}),f(-\tau_{b}))
\end{equation}
where $\simeq$ means equivalent under global $SL(2)$ transformation
(\ref{eq:global-sl2}), namely the $\s$ value is the same. This restricts $v_{2}=v_{1}=v$.
We summarize the translation invariant solution as follows
\begin{align}
f(\tau)=g(\tau)=\f {\tan(\w\tau+v)}{\mJ},~e^{\s}=\f{\w^{2}}{\mJ^{2}\cos^{2}\w\tau_{ab}}\label{eq:mu0-sln}
\end{align}
The UV condition states that when $\tau_{a}=\tau_{b}-\b/2$, we must have
that $e^{\s}=1$. This determines
\begin{equation}
\w=\mJ\cos\w\b/2
\end{equation}

\subsubsection{Twist solution at large $q$} \label{sec:large-q}

\begin{figure}
\begin{centering}
\includegraphics[width=8cm]{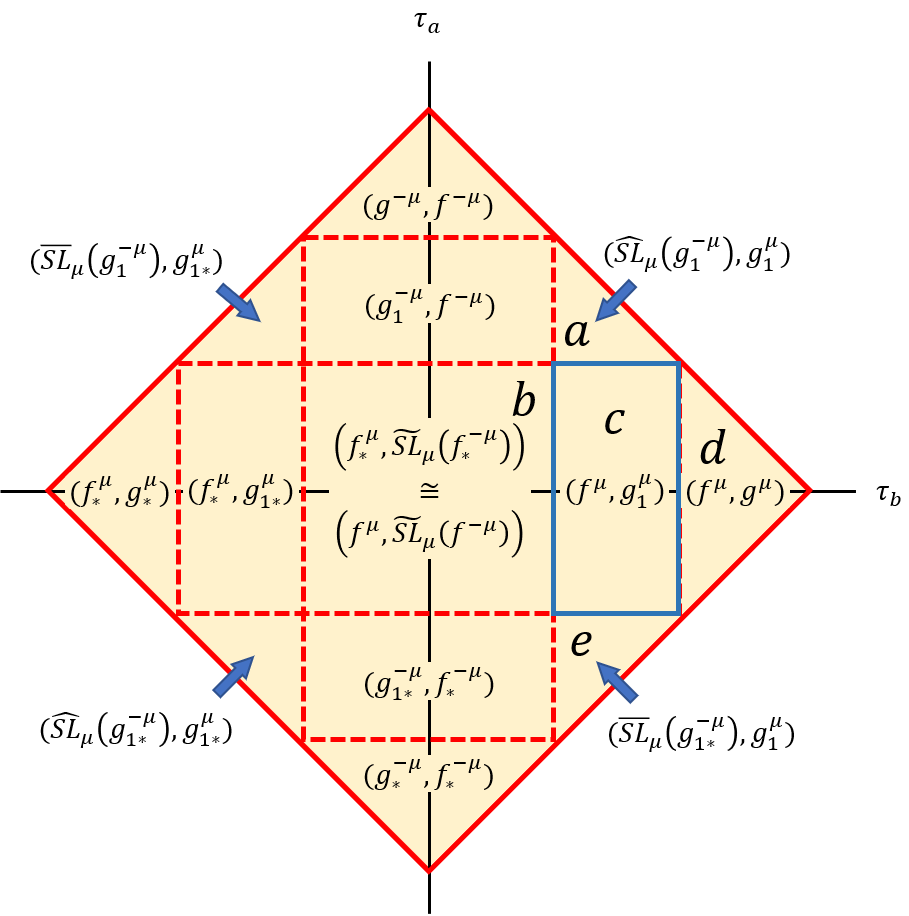}
\par\end{centering}
\caption{The assignment of different $(f_{i},g_{i})$ functions to each region
with all symmetry conditions satisfied in the large $q$ case. We need
to determine three functions $f^{\mu}$, $g_{1}^{\mu}$ and $g^{\mu}$
with twist boundary condition on the blue rectangle, and four $SL(2)$
transformations $\widetilde{SL}_{\mu}$, $\widehat{SL}_{\mu}$, $\overline{SL}_{\mu}$
and $sl_{\mu}$.\label{fig:The-function-}}
\end{figure}

Let us move to nonzero $\mu$ case. The twist boundary condition makes
it hard to find $f$ and $g$ across the various subregions of  the fundamental region. However, as the coupling scales as $1/q$,
the twist boundary condition significantly simplifies in the large $q$ limit.
At order $1/q$, (\ref{eq:tbd-1}) and (\ref{eq:tbd-2}) become
\begin{align}
\s(\tau_{a},(\b/4\mp\e)_{+}) & =\s(\tau_{a},(\b/4\mp\e)_{-})\mp\mu/N\label{eq:bd-1}\\
\s((-\b/4+\e)_{+},\tau_{b}) & =\s((-\b/4+\e)_{-},\tau_{b})-\mu/N\\
\s((\b/4-\e)_{-},\tau_{b}) & =\s((\b/4-\e)_{+},\tau_{b})-\mu/N\label{eq:bd-3}
\end{align}
which means that two twisted components decouple. All of these
conditions can be written in the following form
\begin{equation}
\f{f_{1}'(\tau_{a})g_{1}'(\tau_{b})}{(1+\mJ^{2}f_{1}(\tau_{a})g_{1}(\tau_{b}))^{2}}=\f{e^{\pm\mu/N}f_{2}'(\tau_{a})g_{2}'(\tau_{b})}{(1+\mJ^{2}f_{2}(\tau_{a})g_{2}(\tau_{b}))^{2}}\label{eq:117}
\end{equation}
where one of $\tau_{a,b}$ is fixed to be the value on a border between
subregions, and subscripts 1 and 2 are used to distinguish the functions
$f$ and $g$ on neighboring subregions (for example, 1 could refer
to subregion $d$, 2 could refer to subregion $c$, and we choose
minus sign in (\ref{eq:117})). Let us fix $\tau_{b}$ as in (\ref{eq:117}), and
then we can integrate $\tau_{a}$ to get
\begin{align}
&\f{g_{1}'(\tau_{b})}{\mJ^{2}g_{1}(\tau_{b})(1+\mJ^{2}f_{1}(\tau_{a})g_{1}(\tau_{b}))}=\f{e^{\pm\mu/N}g_{2}'(\tau_{b})}{\mJ^{2}g_{2}(\tau_{b})(1+\mJ^{2}f_{2}(\tau_{a})g_{2}(\tau_{b}))}+c
\end{align}
One can easily find the relation between $f_{1}(\tau_{a})$ and $f_{2}(\tau_{b})$
across the border: they are related to each other via an $SL(2)$ transformation
that depends on the values of $g_{1,2}$ on the border. Note that
this boundary condition does not give any information about how $g_{1}(\tau_{b})$
is related to $g_{2}(\tau_{b})$. Since we have a global $SL(2)$
symmetry, we can do a $SL(2)$ transformation for $f_{2}$ and $g_{2}$
such that $f_{1}=f_{2}$. A similar analysis applies to other borders to fix $b$, $c$ and $d$ with same $f^\mu$ and $a$, $c$ and $e$ with same $g_1^\mu$ (see Fig. \ref{fig:The-function-}).

Note that now the symmetry lines are $\tau_{a}=\pm\tau_{b}$, across
which we have 
\begin{equation}
\mG_{\mu}(\tau_{a},\tau_{b})=\mG_{-\mu}(\tau_{b},\tau_{a}),\;\mG_{\mu}(\tau_{a},\tau_{b})=\mG_{-\mu}(-\tau_{b},-\tau_{a})\label{eq:119}
\end{equation}
Together with the smoothness requirement, this  strongly
constrains the function choice in subregion $b$ (see Fig. \ref{fig:The-function-}).
Suppose the function in $b$ is denoted as $(f_{0}^{\mu},g_{0}^{\mu})$. Then
(\ref{eq:119}) implies that
\begin{equation}
(f_{0}^{\mu},g_{0}^{\mu})\simeq(g_{0}^{-\mu},f_{0}^{-\mu}),\;(f_{0}^{\mu},g_{0}^{\mu})\simeq(g_{0*}^{-\mu},f_{0*}^{-\mu})
\end{equation}
where $*$ denotes the flip $\tau\ra-\tau$. It follows that $g_{0}^{\mu}$
must be a $SL(2)$ transform of $f_{0}^{-\mu}$. Let us assume $
g_{0}^{\mu}=\widetilde{SL}_{\mu}(f_{0}^{-\mu})$
which obeys $(f_{0}^{\mu},\widetilde{SL}_{\mu}(f_{0}^{-\mu}))\simeq(\widetilde{SL}_{-\mu}(f_{0}^{\mu}),f_{0}^{-\mu})$
due to the first symmetry in (\ref{eq:119}). When $\mu=0$, we see from (\ref{eq:mu0-sln})
that the gluing $\widetilde{SL}$ reduces to the identity so that the solution is smooth.

Similarly, $f_{0}^{\mu}$ must be related to $f_{0*}^{\mu}$ by an
$SL(2)$ transformation, as $f_0^\mu=sl_\mu(f_{0*}^\mu)$, due to the second symmetry in \eqref{eq:119} and this $SL(2)$ transformation should also be compatible with $\widetilde{SL}_{\mu}$. This imposes a few constraints on both $SL(2)$ transformations. Such an analysis can be applied to regions $a$ and $e$, in which each have only
one symmetry. Taking all these symmetries in to account, we only need
to determine three functions $f^{\mu}$, $g_{1}^{\mu}$ and $g^{\mu}$
with twist boundary condition on the blue rectangle in Fig. \ref{fig:The-function-}, and four $SL(2)$
transformations $\widetilde{SL}_{\mu}$, $\widehat{SL}_{\mu}$, $\overline{SL}_{\mu}$
and $sl_{\mu}$.

The twist boundary conditions on the four blue segments in Fig. \ref{fig:The-function-}
are given as follows
\begin{align}
g_{1}^{\mu}{'}(\tau_{+}) & =e^{-\mu}g^{\mu}{'}(\tau_{+}),\;g_{1}^{\mu}(\tau_{+})=g^{\mu}(\tau_{+}),\label{eq:b1}\\
\widetilde{SL}_{\mu}(f^{-\mu}){'}(\tau_{-}) & =e^{\mu}g_{1}^{\mu}{'}(\tau_{-}),\;\widetilde{SL}_{\mu}(f^{-\mu})(\tau_{-})=g_{1}^{\mu}(\tau_{-}),\label{eq:b2}\\
\widehat{SL}_{\mu}(g_{1}^{-\mu}){'}(\tau_{-}) & =e^{\mu}f^{\mu}{'}(\tau_{-}),\;\widehat{SL}_{\mu}(g_{1}^{-\mu})(\tau_{-})=f^{\mu}(\tau_{-}),\label{eq:b3}\\
\overline{SL}_{\mu}(g_{1*}^{-\mu}){'}(-\tau_{-}) & =e^{\mu}f^{\mu}{'}(-\tau_{-}),\;\overline{SL}_{\mu}(g_{1*}^{-\mu})(-\tau_{-})=f^{\mu}(-\tau_{-})\label{eq:b4}
\end{align}
where $\tau_{\pm}\equiv\b/4\pm\e$. We can use these equations to show that $\widehat{SL}_{\mu}$ is completely determined by $\widetilde{SL}_{\mu}$. See Appendix \ref{app} for more details.

Let us discuss the UV constraint, which requires that $e^{\s}=1$ when $\tau_{a}=\tau_{b}-\b/2$. In general, this condition is surprisingly easy to satisfy.
In the unshifted coordinate $\tau_{a}$, it means that
\begin{equation}
\f{f'(\tau)g'(\tau)}{(1+\mJ^{2}f(\tau)g(\tau))^{2}}=1\label{eq:137}
\end{equation}
Indeed, for arbitrary $f$ and $g$, we need only to choose
$\tau$ as the proper length in the space $(f,g)\in\R^{2}$ with metric
\begin{equation}
d\tau^{2}=\f{dfdg}{(1+\mJ^{2}f(\tau)g(\tau))^{2}}
\end{equation}
Therefore, in principle we may expect infinitely many solutions for $(f^{\mu},g_{1}^{\mu},g^{\mu})$
satisfying all of the conditions of our problem. % as all analysis above does not constrain the solution too much. 
Indeed, this is because the large $q$ equation of motion \eqref{eq:96} is too loose. The constructive approach would require working to the next order in $1/q$, which should lift the flat direction of solutions. 

However, to get find any solution analytically  is not easy. For example,
it is not clear that after redefining $\tau$ according
to UV requirement,  the functions in region $e$ will be still of the form $(\overline{SL}_{\mu}(g_{1*}^{-\mu}),g_{1}^{\mu})$,
namely that the first component is related to the second one via a reflection
and $SL(2)$. In order to find a reasonably simple solution, we  make further assumptions: 
\begin{equation}
f^{\mu}=f,\;g_{1}^{\mu}=G(f)\equiv\f{A_{\mu}+B_{\mu}f}{C_{\mu}+D_{\mu}f},~\widetilde{SL}_{\mu}=\I \label{eq:139}
\end{equation}
which also implies that $\widehat{SL}_{\mu}=\I$. The first assumption  $f^\mu=f$ is quite strong, but it is a good candidate to satisfy the strong symmetry constraints in subregion $b$. The second assumption is inspired by \eqref{eq:3.14} where moving across $e^{\mu V}$ is a $SL(2)$ transformation for $\psi$. 

An important consistency criterion on the solution is that when it is continued to Lorentz signature and the limit $\e \rightarrow 0$ is taken, it should obey causality. In other words, the correlation functions in the past of the interaction $e^{i g V}$ must be the original ones of the thermofield double state in the decoupled system. We will see later that our solution has this property. For example in the region b, with our ansatz $\avg{\psi_l(t')e^{igV}e^{-igV}\psi_r(t)}$ is the original translation invariant two point function. However in region d, only after taking $\e \ra 0$ can one see that the Lorentzian solution obeys causality. 

With this ansatz, solving for $(f_i,g_i)$ in all subregions is straightforward. We leave the details to Appendix \ref{app} and only describe the result here. Taking $\tau_{+}=\tau_{-}=\b/4$
and shifting $\tau_{a}$ back $\tau_{a}\ra\tau_{a}-\b/2$, the solutions of $e^{\s}$ in each subregion are as follows:
\begin{align}
\left[e^{\s}\right]_{(a)} =&\f{e^{-2\mu/N}\w^{2}}{\mJ^{2}}\left[\cos\w(\tau_{ab}-\f{\b}2)+\f{(e^{-\mu/N}-1)}{\cos\w\b/2}\cos\w\tau_{ab}\right.\nn\\
&\left.+\f{(e^{-\mu/N}-1)^{2}\sin\w\b/2}{\cos^{2}\w\b/2}\cos\w(\tau_{a}-\f{\b}4)\sin\w(\tau_{b}-\f{\b}4)\right]^{-2}\label{eq:156}\\
\left[e^{\s}\right]_{(b)} =&\left[e^{\s}\right]_{(e)}=\left[e^{\s}\right]_{(d)}=\f{\w^{2}}{\mJ^{2}\cos^{2}\w(\tau_{ab}-\f{\b}2)}\\
\left[e^{\s}\right]_{(c)} =&\f{e^{-\mu/N}\w^{2}}{\mJ^{2}\left(\cos\w(\tau_{ab}-\f{\b}2)+\f{(e^{-\mu/N}-1)}{\cos\w\b/2}\sin\w(\tau_{a}-\f{\b}4)\sin\w(\tau_{b}-\f{\b}4)\right)^{2}}
\end{align}
It is interesting that the solution in subregion $c$ has the same form as that in \cite{qi2018quantum}.

\section{Effectiveness of protocol} \label{sec:effect}

In order to check the mutual information $I_{RT}$ after performing the protocol, we need to evaluate correlation functions in Lorentz signature. The Lorentzian correlation functions are analytic continuations of
$\mG_{\mu}$. For example, the OTOC $\avg{\mU^\dagger\psi_{l}^{1}(t)\mU\psi_{l}^{1}(t)}$ is
equal to
\begin{equation}
\lim_{\e\ra0}\mG_{i\mu}(\b/4+it,(\b/4-\e)_{-}+it)
\end{equation}

\subsection{Lorentzian correlation function}

To find the density matrix $\r_{ij}$, we need to calculate a few OTOCs. We first consider
\begin{align}
\mK&(t,t')=\avg{\{\psi_{l}(-t'),\mU^\dagger\psi_{r}(t)\mU\}}=\avg{\psi_{l}(-t')\mU^\dagger\psi_{r}(t)\mU}+\avg{\mU^\dagger\psi_{r}(t)\mU\psi_{l}(-t')}
\end{align}
Comparing with the definition of $\mG_{\mu}$, we can easily find
that
\begin{align}
\mK(t,t')& =-i\lim_{\e\ra0}\left[\mG_{i\mu}(-it'+\b(1/4+\e)_{+},3\b/4-it)+\mG_{i\mu}(3\b/4-it,-it'+\b(1/4-\e)_{-})\right]\nonumber \\
 & =\f i2\left(\f{e^{-i\mu/(Nq)}\w^{2/q}}{\mJ^{2/q}\left[\cosh\w\D t-\f{\mJ(e^{-i\mu/N}-1)}{\w}\sinh\w(t-i\b/2)\sinh\w t'\right]^{2/q}}-h.c.\right)\label{eq:K}
\end{align}
where we used symmetries from Sec. \ref{subsec:Symmetries} to express $\mG_\mu$ in the fundamental domain and in the last line $\D t\equiv t'-t$.

There is another type of OTOC involved in $\r_{ij}$, namely $\avg{\psi^{1}\psi^{2}\mU^\dagger\psi^{1}\psi^{2}\mU}$.
One can follow a similar procedure as in the calculation of $\mG$ to derive the resulting twist boundary
condition for four point correlation functions
\begin{equation}
\hat{\mG}(\tau_{a},\tau_{b};\tau_{c},\tau_{d})=\f{\avg{I|e^{-\b H/2}\bar{\mT}\left[e^{-\mu V(\b/4+\e)}e^{\mu V(\b/4-\e)}\psi^{1}(\tau_{a})\psi^{1}(\tau_{b})\psi^{2}(\tau_{c})\psi^{2}(\tau_{d})\right]|I}}{\avg{I|e^{-\b H/2}e^{-\mu V(\b/4+\e)}e^{\mu V(\b/4-\e)}|I}}\sgn(\bar{\tau}_{a},\bar{\tau}_{b},\bar{\tau}_{c},\bar{\tau}_{d})
\end{equation}
where $\sgn(\bar{\tau}_{a},\bar{\tau}_{b},\bar{\tau}_{c},\bar{\tau}_{d})=1$
for $\bar{\tau}_{a}>\bar{\tau}_{b}>\bar{\tau}_{c}>\bar{\tau}_{d}$
and the sign flips for every permutation of the order. It is very straightforward
to see that for fixed $\tau_{c}$ and $\tau_{d}$, the correlator as a function of $\tau_{a}$ and
$\tau_{b}$ obeys the same twist boundary condition when crossing $e^{\pm\mu V}$ as before.
The same applies to the boundary condition for $\tau_{c}$ and $\tau_{d}$ with
$\tau_{a}$ and $\tau_{b}$ fixed. Therefore, we can understand
$\hat{\mG}$ as a four point function $\avg{I|e^{-\b H/2}\bar{\mT}\left[\psi^{1}(\tau_{a})\psi^{1}(\tau_{b})\psi^{2}(\tau_{c})\psi^{2}(\tau_{d})\right]|I}$
with two copies of the twist boundary conditions. On the other hand, due
to $SO(N)$ global symmetry and large $N$ limit
\begin{align}
 & \avg{I|e^{-\b H/2}\bar{\mT}\left[G(\tau_{a},\tau_{b})G(\tau_{c},\tau_{d})\right]|I}\nonumber \\
= & \avg{I|e^{-\b H/2}\bar{\mT}\left[\psi^{1}(\tau_{a})\psi^{1}(\tau_{b})\psi^{2}(\tau_{c})\psi^{2}(\tau_{d})\right]|I}+O(N^{-1})\label{eq:144}
\end{align}
where we suppressed the insertion of $e^{\pm\mu V}$ for notational
simplicity. In the large $N$ limit, the leading order contribution to $\avg{I|e^{-\b H/2}\bar{\mT}\left[G(\tau_{a},\tau_{b})G(\tau_{c},\tau_{d})\right]|I}$
is disconnected, namely the product of two correlation functions $G(\tau_{a},\tau_{b})G(\tau_{c},\tau_{d})$.

\begin{figure}
\begin{centering}
\includegraphics[width=7cm]{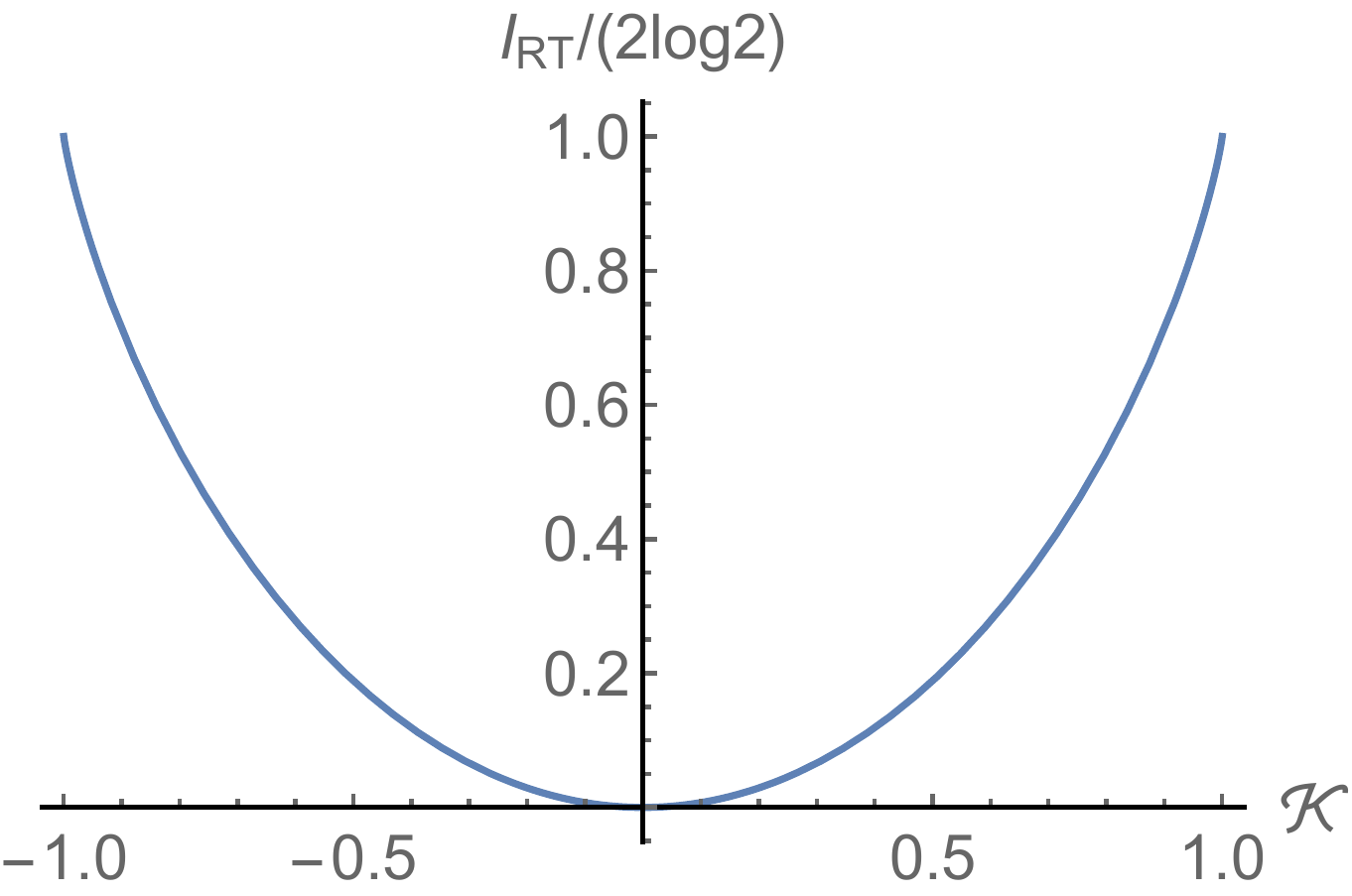}
\par\end{centering}

\caption{The plot of $I_{RT}/(2\log2)$ as a function of $\protect\mK$. $I_{RT}$
reaches its maximal value only when $\protect\mK=\pm1$. \label{fig:3a}}
\end{figure}

Now we are readily write down the contribution to $\r_{ij}$ from
this type of OTOC. In $\r_{11}$, we have 
\begin{align}
 & \avg{\{\psi_{l}^{1}(-t'),[\psi_{l}^{2}(-t'),\mU^\dagger\psi_{r}^{1}(t)\psi_{r}^{2}(t)\mU]\}}\nonumber \\
\app & \f 1{N^2}\sum_{i,j}\avg{\{\psi_{l}^{i}(-t'),[\psi_{l}^{j}(-t'),\mU^\dagger\psi_{r}^{i}(t)\psi_{r}^{j}(t)\mU]\}}\nonumber \\
\app & -\avg{\{\psi_{l}^{1}(-t'),\mU^\dagger\psi_{r}^{1}(t)\mU\}}\avg{\{\psi_{l}^{1}(-t'),\mU^\dagger\psi_{r}^{1}(t)\mU\}}\nonumber \\
\app & -\mK(t,t')^{2}
\end{align}
where in second line we used the identity (\ref{eq:144}) and the fact that $\psi^1_r(t)\psi^1_r(t)=1/2$, and in third line we used factorization at large $N$. Similarly, the correlation function appearing in 
 $\r_{14}$ is given by
\begin{align}
\avg{\{\psi_{l}^{2}(-t'),\psi_{l}^{1}(-t')\mU^\dagger\psi_{r}^{2}(t)\mU\psi_{l}^{1}(-t')\}} \app & -\f 12\mK(t,t')
\end{align}
From these results, we obtain
\begin{align}
\r_{11}=\f 12(1+\mK(t,t')^{2}),\;\r_{14}=\mK(t,t')
\end{align}
and other $\r_{ij}$ via \eqref{eq:46-0} to \eqref{eq:46}. From (\ref{eq:K}) we know that $\mK$ (short for $\mK(t,t')$) is a real valued function, which
is consistent with the fact that $\r$ is hermitian. It follows that
the mutual information between $R$ and $T$ is 
\begin{align}
I_{RT}=&\f 14\left[(\mK-1)^{2}\log(\mK-1)^{2}+(\mK+1)^{2}\log(\mK+1)^{2}+2(1-\mK^{2})\log(1-\mK^{2})\right]\label{eq:148}
\end{align}
We plot the dependence of $\mK$ in Fig. \ref{fig:3a}. We see that
$I_{RT}$ is close to maximal when $\mK$ approaches to $\pm1$. 

\subsection{Semiclassical limit}

%\DLJ{Our entire analysis assumes large $N$, otherwise we wouldn't reduce to solving classical equations for $G$, etc. I don't fully understand what you meant in the next paragraph. Also, I think this subsection should be renamed - part of it is about comparison to the gravity result, and part is putting everything together in our setup.}
%There is a large $N$ limit in the left-right coupling. Recall that from (\ref{eq:36}) we couple $\psi_{l}^{i}$ with $\psi_{r}^{i}$ for all $i$ with coefficient $\mu/(qN)$. As here $1/N$ appears, we need to clarify what it means to have a large $q$ limit solution though $q\ll N$. Indeed, we can interpret the result as a large $q$ expansion but at each order we consider the effect up to $1/N$. In other words, this is re-organization of large parameter expansion. On the other hand, this $1/N$ effect comes from the coupling coefficient and directly from the twist boundary condition. In previous studies, $1/N$ effect comes from Schwarzian effective action. It seems that the twist boundary condition automatically takes care of the Schwarzian and resummed the $1/N$ effect into a closed form. \cite{qi2018quantum} also shows this feature.

%\DLJ{Here's my understanding, do you agree?} \PG{I agree. And I think your version sounds better. I would prefer to call the name of this subsection as "Semiclassical limit".}

Our analysis relied on the leading $1/N$ description of the SYK model in terms of classical $G$ and $\Sigma$ collective fields. The interaction (\ref{eq:36}) coupled all $N$ fermions $\psi_{l}^{i}$ with $\psi_{r}^{i}$ with a coefficient which we took to scale as $\mu/(qN)$ for fixed $\mu$. 
We further specialized to the large $q$ limit, although $q\ll N$. In other words, at each order in the $1/q$ expansion we kept only the leading effects in $1/N$.

Due to the instantaneous nature of the interaction, we treated it exactly in terms of the fundamental Majoranas, which resulted in particular twist boundary conditions for the collective fields across the interaction time. In this way, the $1/N$ effect arising from the Schwarzian effective action that dominates at large $\beta \mJ$ is automatically resummed into a closed form by the full solution with twist boundary conditions, as in 
\cite{qi2018quantum}. Moreover, our analysis includes effects beyond the Schwarzian approximation, which are important for understanding how the operations involving the fundamental Majoranas glue into the gravitational dual near the boundary, as discussed below. We will see that those ``stringy'' corrections are crucial for regulating the pole in the gravity transmission amplitude in the probe approximation and are thus very important for finding the effectiveness of the teleportation protocol. It will not require going to higher orders in the $1/N$ expansion.

%\DLJ{Is part of the point here that in MSY, the pole is cut off by higher $1/N$ corrections, ie. from resumming the Schwarzian, while for us, at large $q$, the pole is cut off by ``stringy'' corrections even at the leading order in $1/N$?} \PG{I agree that in MSY, the cut off of pole comes from higher order of $1/N$, namely $G_N$. However, I do not think the resummation of Schwarzian will give full $1/N$ scattering result. You can see from (6.54) of \cite{maldacena2016conformal}, there they did not resolve the pole. I am not quite sure how to use Schwarzian to completely recover the scattering result in MSY. For us, it is true that in leading order of $1/N$, we get ``stringy" correction that cut off the pole.}

The scrambling time required for traversability is associated to the limit of large  $N$ with $e^{\w(t+t')}/N$ fixed. Then $\mK$
becomes
\begin{equation}
\mK=-\Im\left(\f{\w^{2/q}}{\mJ^{2/q}\left[\cosh\w\D t+i\f{\mJ\mu}{4\w N}e^{\w(t+t')}e^{-i\w\b/2}\right]^{2/q}}\right)
\end{equation}
In the strong coupling or low temperature limit, $\b\mJ\gg1$, the frequency becomes
\begin{equation}
\w=\mJ\cos\w\b/2\implies\w=\f{\pi}{\b}-\f{2\pi}{\b^{2}\mJ}+O(1/\mJ^{2})
\end{equation}
Plugging this into $\mK$, we get
\begin{equation}
\mK\app -\Im \left( \left[\f{\pi/(\b\mJ)}{\cosh\pi\D t/\b+\f{\mJ\mu\b}{4\pi N}e^{\pi(t+t')/\b}e^{i\pi/(\b\mJ)}}\right]^{2/q}\right)\label{eq:168}
\end{equation}
which has the maximal Lyapunov exponent of $2\pi/\b$ \cite{Maldacena:2015waa}. 

A quick conclusion from \eqref{eq:168} is that ignoring the numerator, if $\mJ,N/\mJ\ra\infty$, a
negative $\mu$ will lead to a pole for large enough $t+t'$. To compare with gravity result explicitly, we can define
\be
g=-\mu,\quad \D=1/q,\quad G_N=\f{4^\D\b\mJ}{\pi\D N}
\ee
and the first term in \eqref{eq:168} now becomes
\be 
\left( \f{\pi}{\b\mJ}\right)^{2\D}\f{1}{\left(\cosh\pi\D t/\b-\f{\D g}{2^{2\D+2}} G_N e^{\pi(t+t')/\b}\right)^{2\D}}
\ee
which exactly matches with the semiclassical gravity result of \cite{maldacena2017diving} (associated to traversal of a probe signal, ignoring its backreaction). Such a pole indicates that the bulk geometry is a traversable wormhole in which there exist timelike geodesics connecting the $l$ system at $-t'$ with the $r$ system at $t$. The negative sign required  for $\mu$ is also consistent with \cite{maldacena2018eternal}, in which a large coupling $\mu$ that remains on for all time will result in thermofield double state becoming an eternal traversable wormhole. 

\begin{table}
\begin{centering}
\begin{tabular}{|c|c|c|c|c|c|}
\hline 
$q$ & 4 & 8 & 12 & 16 & 20\tabularnewline
\hline 
$\max(I_{RT})/(2\log2)$ & 0.540 & 0.201 & 0.105 & 0.066 & 0.045\tabularnewline
\hline 
\end{tabular}
\par\end{centering}
\caption{Maximal $I_{RT}$ for first a few $q$'s. \label{tab:Maximal--for}}
\end{table}

An actual pole in the result would violate unitarity, since the fermion number operator must have an expectation value between 0 and 1. This is consistent with our result, as due to the existence of numerator, the value of $\mK$ does not blow up. This comes from the UV physics of SYK model and cannot be seen in the low energy gravity theory. In the large $\mJ$ limit, the numerator $1/(\b\mJ)$ and the exponential $e^{i\pi/(\b\mJ)}$ perfectly balance to give a finite maximal value of $|\mK|$. 

This $1/(\b\mJ)$ effect comes from a correction to maximal Lyapunov exponent, thus we may refer it as a ``stringy'' effect in the putative dual bulk theory of SYK. For evaluating the effectiveness of the teleportation protocol, this stringy effect is crucial, as it determines the actual maximal value of $|\mK|$. The nontrivial exponential and the $1/\b\mJ$ correction to Lyapunov exponent are also related to the near coherent scrambling discussed in \cite{kitaev2017near}.

At weak coupling or high temperature, $\b \mJ\ll 1$, the frequency is given by $\w\app \mJ$, and $\mK$ becomes
\begin{equation} \label{eq:highT}
\mK\app -\Im\left(\left[\f 1{\cosh\mJ\D t+\f{i\mu}{4N}e^{\mJ (t+t')}}\right]^{2/q}\right)
\end{equation}
As $t+t'$ grows, $\mK$ also has a peak, but it is not as high and sharp as in the low temperature case. In particular, the denominator then never leads to a pole. This is unsurprising because at high temperatures, the full stringy dual of the SYK model is unknown, and is certainly not semi-classical gravity. It is also worth noting that the Lyapunov exponent now is $2\mJ$, as expected \cite{roberts2018operator}. In this paper we are mainly interested in the low temperature limit and leave discussion of high temperature to Sec. \ref{sec:inter}.

To see the location and value of the peak of $\mK$ in \eqref{eq:168}, it is useful to define parameters
\begin{equation}\label{eq:4.16}
a\equiv\pi/(\b\mJ),\;x\equiv \f{-\mu}{4N} e^{2\pi t/\b},\;y\equiv \f{-\mu}{4N} e^{2\pi t'/\b}
\end{equation}
in terms of which (\ref{eq:168}) can be written as
\begin{align}
\mK=&-\left[ \f{4a^4xy}{4x^2y^2+a^2(x+y)^2-4axy(x+y)\cos a}\right]^{1/q}\nn\\
&\times\sin \left[ \f 2 q \arccos\f{a (x + y) - 2 x y \cos a}{\sqrt{4x^2y^2+a^2(x+y)^2-4axy(x+y)\cos a}} \right]
\end{align}
To find the minimal value of $\mK$ (note that $\mK<0$), we extremize it with respect to $x$ and $y$. Expanding in small $a$, the solution is
\be\label{eq:4.18}
x=y=a+a^2 \cot \f{2\pi}{q+2}+O(a^3)
\ee
which defines a time $t=t'=t_0$ by \eqref{eq:4.16}. Using this solution, we find that the minimal value of $\mK$ around $a\app 0$ is achieved when $\mJ\ra\infty$, $a\ra 0$. In this limit, the minimal value of $\mK$ is
\be\label{eq:4.19}
\mK_{\min} = -\left[\sin \f {2\pi}{q+2}\right]^{1+2/q}
\ee
From this expression and \eqref{eq:148}, we see that $|\mK_\min|$ and $\max(I_{RT})$ decrease as $q$ becomes larger. This is reasonable as $\mK$ is the imaginary part of the first term in \eqref{eq:168} and $q$ controls its phase, which decreases as $1/q$. For large $q$ the scaling is 
\be 
\mK_\min \sim -2\pi/q,\; \f{\max(I_{RT})}{2\log 2}\sim \f{2 \pi^2}{ q^2 \log 2}
\ee
We present the  maximal values of $I_{RT}$ for a several  values of $q$ in Table \ref{tab:Maximal--for}.

With this protocol, the large $q$ limit does not lead to perfect teleportation. However, the mutual information does grow in time from  exponentially small values at early times to an order
1 number after the scrambling time. By examining the density matrix in more detail one can determine whether quantum entanglement between $R$ and $T$ is generated, or if only classical correlation appears. In Sec. \ref{sec:classical} we show that the resulting density matrix is inseparable.  Thus we
could then imagine taking an $O(1)$ number of copies of SYK model and performing 
entanglement distillation \cite{bennett1996concentrating, bennett1996mixed, bennett1996purification}, %. This is sort of standard procedure in quantum
%information that can 
which would increase the fidelity of teleportation. %However, entanglement distillation requires local operations and classical communications on both sides. 

It turns out that there is a more elegant way of improving the protocol. In Sec. \ref{sec:improve} we shown that applying the initial SWAP operation to a product of order $q$ fermions  achieves almost fidelity of teleportion.

\begin{figure*}
\begin{centering}
\subfloat[$t'<t_{s}$\label{fig:5a}]{\begin{centering}
\includegraphics[height=3.8cm]{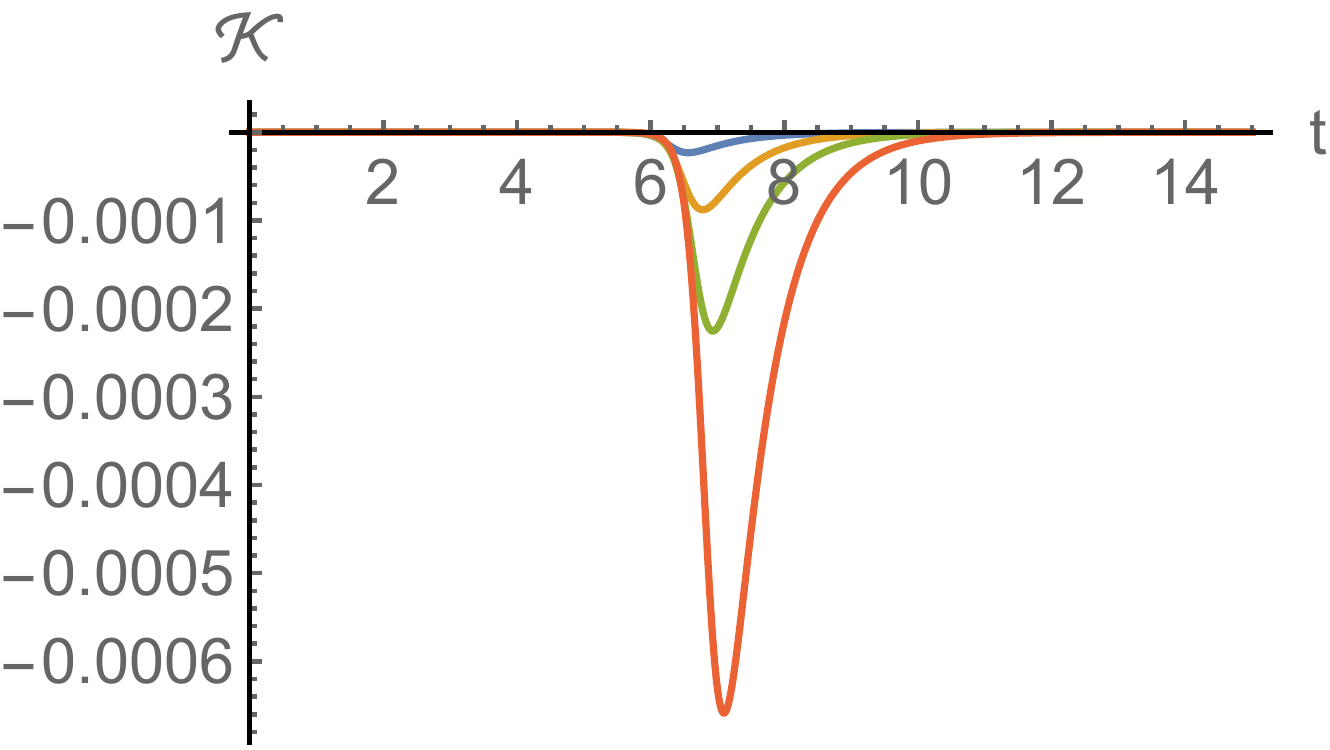}
\par\end{centering}
}\subfloat[$t'>t_{s}$\label{fig:5b}]{\begin{centering}
\includegraphics[height=3.8cm]{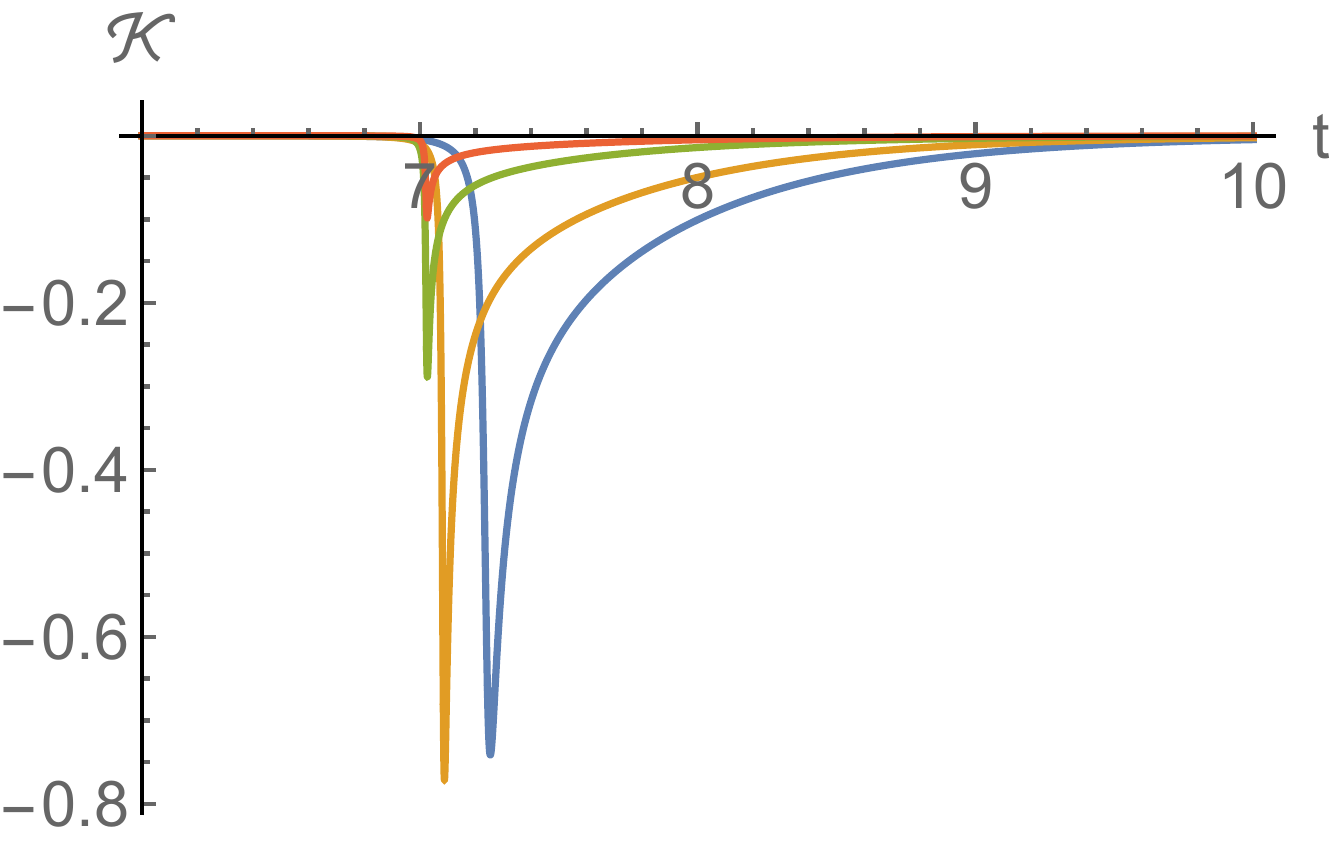}
\par\end{centering}
}\\
\subfloat[\label{fig:5c}]{\begin{centering}
\includegraphics[height=3.8cm]{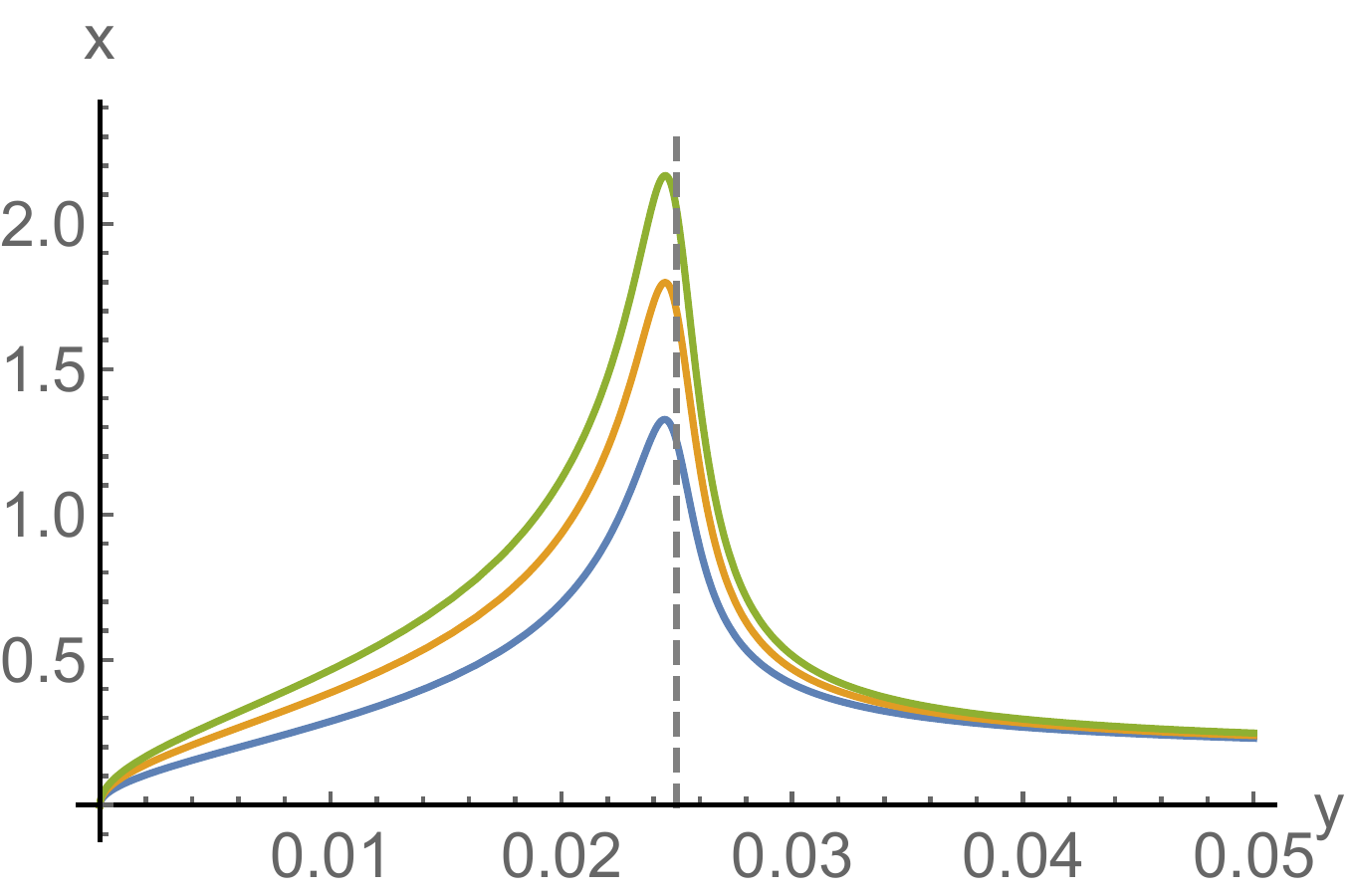}
\par\end{centering}
}
\par\end{centering}
\caption{The plot of $\protect\mK$ as a function of arrival time $t$ for
different injection time $-t'$ around transition time $-t_{s}$. Blue,
yellow, green and red correspond to increasing $t'$ (earlier injection).
(a) shows $t'<t_{s}$ where peaks move to right and have opposite
ordering of signal, and (b) shows $t'>t_{s}$, where peaks move to
left and have same ordering of signal. (c) shows the peak location
$x(y)$ for $q=4,8,12$ (blue, yellow and green). It is clear that
there exists transition point $t_{s}$ at which the time ordering
of signal flips. The dashed line is $y=a/2$. \label{fig:5}}
\end{figure*}

We plot various features of the time dependence of the transmitted signal in Fig. \ref{fig:5}. The final signal, $\mK$, is plotted  as a function
of time, $t$, for several different initial injection times $-t'$ into the left SYK system in Fig. \ref{fig:5}. Characterizing the arrival time of the signal by the location
of peak, we see there are two distinct cases.
If the injection time is not sufficiently early (in Fig. \ref{fig:5a}),
the peaks move to later times when we have earlier injection. In other words, the time ordering of the signal
 is inverted. On the other hand, if the injection time is early
enough (in Fig. \ref{fig:5b}), the peaks move to earlier times when we have
earlier injection. That means that the time ordering of the signal is preserved.

The latter behavior is what is expected for travel through a traversable wormhole in semiclassical gravity. Denote the transition time between the two behaviors as $t' = t_s$. 
From Fig. \ref{fig:5a} and Fig. \ref{fig:5b}, it is clear that the
signal significantly strengthens when $t'>t_{s}$. The range
$t'>t_{s}$ should be thought of as the semiclassically traversable regime. Outside of this semiclassical regime, the time ordering is reversed due to the pattern of entanglement in the thermofield double state, which is consistent with the analysis of \cite{gao2018regenesis}.

We can also find the peak arrival time, $t$ for which $\del_{t}\mK=0$, as a function of the injection time $-t'$. % solve the function $t(t')$ directly for $\del_{t}\mK=0$
%to find the location of the peak and 
  The arrival time $t(t')$ achieves its maximum
when $t'=t_{s}$. As an example, for large $\mJ$ we plot this function in terms of exponentiated variables $x(y)$ for various values of $q$ in Fig. \ref{fig:5c}. It is
clear that there exists a transition point $t_{s}$ where the time
ordering of the signal flips. For small $a$ (large $\mJ$), this
transition point in terms of the variables $x$ and $y$ can be calculated as a
series in $a$:
\begin{equation}
\begin{aligned}\label{eq:4.21}
x_{s}&=\sqrt{\f{q+2}q}\f{\sin^{2}q\theta/2}{\sin\theta}+O(a),\\
y_{s}&=\f 12a-\f{a^{2}}{2\sqrt{q(q+2)}\sin\theta}+O(a^{3})
\end{aligned}
\end{equation}
where $\theta$ is the first positive solution to the transcendental
equation
\begin{equation}
\tan\f{q\theta}2=\f{\sqrt{q}\sin\theta}{\sqrt{q+2}-\sqrt{q}\cos\t}
\end{equation}
whose solution at large $q$ is $\theta\app\f{2.331}{q}.$ The
transition time $t_{s}$ is defined in terms of $y_{s}$ by 
\begin{equation}
t_{s}\equiv T_{0}+\f{\b}{2\pi}\log y_{s},\quad T_{0}=\f{\b}{2\pi}\log\f{4N}{-\mu}
\end{equation}
which is the latest injection time $-t'=-t_{s}$ for which the signal
preserves time ordering. Similarly, $t_{\max}$ is defined via $x_{s}$
by
\begin{equation}
t_{\max}\equiv T_{0}+\f{\b}{2\pi}\log x_{s}
\end{equation}
which is the latest arrival time $t=t_{\max}$ after which $\mK$ has no peak. 

Comparing \eqref{eq:4.18} with \eqref{eq:4.21}, we find that the minimal value of $\mK$ occurs with injection time $-t_0$, which is before the transition time. This matches our observation in Fig. \ref{fig:5}, and $\mK_\min$ in \eqref{eq:4.19} is in the semiclassical regime. Moreover, we can plug the value of transition time \eqref{eq:4.21} into $\mK$ and find $\mK \sim a^{1/q}$ in the small $a$ limit. This means that little signal gets through if injected at or after the transition time, in the low
temperature limit.  Note that $a\sim 1/\b\mJ$ controls ``stringy" effects in the bulk. Physically, injections later than transition time with inverse timing ordering classically do not get through the wormhole at all and would die in the singularity. However, injections close to the transition time travel close enough to the (ultimate) horizon that fluctuations in the effective metric determine whether they get through. Such fluctuations are due to ``stringy" effect as indicated by the $|\mK|$ peak value being suppressed
by $(\b\mJ)^{-1/q}$.

%Comparing (4.18) with (4.21), we find find that the minimal value of K occurs with injection time −t0, which is before the transition time. This matches our observation in Fig. 4.2, and Kmin in (4.19) is in the
%semiclassical regime. Moreover, we can plug the value of the transition time (4.21) into K and find K ∼ a1/q
%in small the a limit. This means that little signal gets through if injected at or after the transition time, in the low
%temperature limit. Note that a ∼ 1/βJ controls “stringy” effect in the dual bulk. Physically, injections
%later than transition time with inverse timing ordering classically do not get through
%the wormhole at all and would die in the singularity. However, injections close to the transition time travel close enough to the (ultimate) horizon that fluctuations in the effective metric
%determine whether they get through. Such fluctuations are due to “stringy” effects as indicated by the |K| peak value being suppressed
%by (βJ )−1/q .

\begin{figure}
\begin{centering}
\includegraphics[width=5cm]{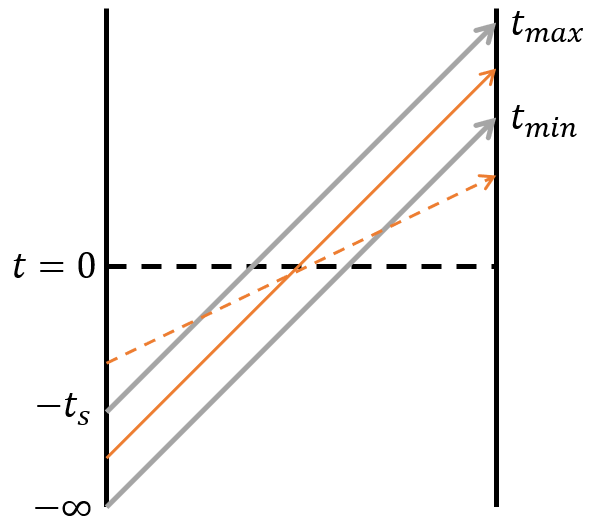}
\par\end{centering}

\caption{Three time scales of $\mK$. We define the arrival time as the peak of $\mK$. The signal emitted from $-t_s$ arrives at maximal time $t_{\text{max}}$ and that from $-\infty$ arrives at minimal time $t_{\text{min}}$. We define $t'>t_s$ as semiclassical regime because the ordering of signal is preserved (solid orange arrow). For $t'<t_s$, the ordering of signal is reversed (dashed orange arrow). \label{fig:12}}
\end{figure}

There is another time scale appears Fig. \ref{fig:5c} in the limit $y\ra\infty$, so the qubit is injected very early. It corresponds
to the lower bound $x_{\min}$ of $x$ that remains in the semiclassical regime. we calculate this as a series in $a$ by taking the equation $\del_{t}\mK=0$
in the $t'\ra\infty$ limit
\begin{equation}
x_{b}=\f 12a+\f 12a^{2}\cot\f{2\pi}{q+2}+O(a^{3})
\end{equation}
This  defines the earliest semiclassical arrival time 
\begin{equation}
t_{\min}\equiv T_{0}+\f{\b}{2\pi}\log x_{b}
\end{equation}
The physical meaning of $t_{\min}$ is obvious. Any signal arriving  semiclassically via the
traversable wormhole cannot arrive earlier than this time. It is reasonable
to define the apparent duration $d$ of the throat of the traversable wormhole, as measured in boundary time on the right, by
\begin{equation}
d=t_{\max}-t_{\min}\app\f{\b}{2\pi}\log\b\mJ
\end{equation}
It is interesting that the duration of the wormhole does not depends on $\mu$ and becomes
divergent in the large $\mJ$ limit. These time scales and the two regimes of time orderings of the signal are shown in Fig. \ref{fig:12}.

After the peak, Fig. \ref{fig:5a} and Fig. \ref{fig:5b} show that $\mK$ has an exponential decay. This decay rate is easily seen from \eqref{eq:168}
\be
\mK\sim e^{-\D(2\pi t/\b)}~\text{for $t$ large}
\ee
where $\D=1/q$ is the conformal weight of the Majorana fermion $\psi$. Thus the signal experiences a thermalization process after transmission to the $r$ system.

We make one final consistency check of our solution in 
the large $N$ limit. In (\ref{eq:156}),
if we take $\tau_{a}\ra3\b/4-it'$ and $\tau_{b}\ra\b/4+it$, we get
\begin{align}
e^{\s}=&\f{\w^{2}}{\mJ^{2}}\left[\cosh\w\D t-\f{i\mu}N(\cosh\w\D t+i\tan\w\b/2\sinh\w\D t)\right.\nn\\
&\left.-\f{\mu^{2}\tan\w\b/2}{N^{2}}(i\cosh\w t'-\tan\w\b/2\sinh\w t')\sinh\w t\right]^{-2}\label{eq:170}
\end{align}
which corresponds to the real correlation function
\begin{equation}
\avg{\text{tfd}|\mU^\dagger i\psi_{r}(t')\psi_{l}(t)\mU|\text{tfd}}
\end{equation}
 One can see that the denominator in (\ref{eq:170}) is real
when $\D t$ and $e^{\omega (t + t'}/N$ are order 1, so that it simply reduces to the first term at leading order in the $1/N$ expansion. The imaginary part that appears at higher order cannot be trusted, as  then 
loop corrections must also be included. 

\subsection{Improved protocol with multiple fermion SWAP} \label{sec:improve}

We found in \eqref{eq:4.19} that $|\mK_\min|$ monotonically decreases  with increasing $q>2$. It is natural to expect that by encoding the teleported qubit in multiple fermions, the fidelity of the protocol can be improved. Noting that the $q$ dependence of the exponent  in \eqref{eq:168} comes from the $1/q$ conformal weight of $\psi$, we will consider making use of heavier operators made out of a product of several fermions.

To be more precise, we consider the following operators in each SYK system, where $p$ is an odd number.
\begin{align}
\Psi^1_{l,r}&=2^{(p-1)/2}i^{p(p-1)/2}\psi^1_{l,r} \psi^3_{l,r} \cdots \psi^{2p-1}_{l,r},\\ \Psi^2_{l,r}&=2^{(p-1)/2}i^{p(p-1)/2}\psi^2_{l,r} \psi^4_{l,r} \cdots \psi^{2p}_{l,r}
\end{align}
which satisfy
\be 
\{\Psi^{i}_{a},\Psi^{j}_{b}\}=\d^{ij}\d_{ab},\quad i,j=1,2,\;a,b=l,r
\ee
Therefore, we can use $\Psi^i$ to construct Dirac fermions $\chi$ and $\chi^\dagger$ as in Sec. \ref{sec:setup}
\be
\chi_{l,r}=\f 1 {\sqrt{2}}(\Psi^1_{l,r}+i\Psi^2_{l,r}),\;\chi_{l,r}^\dagger=\f 1 {\sqrt{2}}(\Psi^1_{l,r}-i\Psi^2_{l,r})
\ee
which implies $\{\chi_{l,r},\chi_{l,r}^\dagger\}=1$. As the $\Psi^i$ obey exactly the same algebra as a single Majorana fermion when acting on themselves, the density matrix elements are the same as \eqref{eq:46-0}-\eqref{eq:46} with the replacement of  $\psi$ by $\Psi$. 

We take $p$ to be a fixed number that does not scale with $N$, so that the correlation functions involving $\Psi^i$ will factorize into  products of correlations functions of the component Majonana fermions (with the usual twist boundary conditions for each component). It follows that the mutual information formula \eqref{eq:148} for $I_{RT}$ still holds but with $\mK\ra\mK_p$ given by 
\begin{align} 
\mK_p=\avg{\{\Psi_{l}(-t'),\mU^\dagger\Psi_{r}(t)\mU\}}=2^{p-1}\left[\avg{\psi_{l}(-t')\mU^\dagger\psi_{r}(t)\mU}^p+\avg{\mU^\dagger\psi_{r}(t)\mU\psi_{l}(-t')}^p\right]
\end{align}
where we omitted the superscript of $\Psi$. For low temperatures, its value is
\be 
\mK_p\app i^{p+1}\Im\left(\left[\f{\pi/(\b\mJ)}{\cosh\pi\D t/\b+\f{\mJ\mu\b}{4\pi N}e^{\pi(t+t')/\b}e^{i\pi/(\b\mJ)}}\right]^{2p/q}\right)
\ee
where $i^p=\pm i$ is pure imaginary as $p$ is an odd number. All of our previous analysis applies with a rescaled $q\ra q/p$. This is reasonable as $\Psi^i$ consists of $p$ Majorana fermions and thus has conformal weight $p/q$.

We are interested in the $p$ value that leads to the highest fidelity teleportation. Odd $p$ guarantees  that (4.35) is pure imaginary, so this means maximizing  $|\mK_p|$. Using \eqref{eq:4.19} we have
\be \label{eq:maxKp}
|\mK_{p}|_\max = \left[\sin \f {2\pi}{q/p+2}\right]^{1+2p/q}
\ee
where it reaches the maximum value of $1$  at $p/q=2$. This is perfect fidelity teleportation. However, as we assumed that $q\in 4\Z$ and $p$ is odd, we can only choose $p=q/2 \pm 1$. Of these, $p=q/2+1$ will give the larger $|\mK_{p}|_\max$. Therefore the improved protocol gives
\be 
|\mK|_{\max}\equiv |\mK_{q/2+1}|_\max= \left[\sin \f {(q+2)\pi}{2(q+1)}\right]^{2+2/q}
\ee
which approaches $1$ when $q \ra \infty$. Expanding at large $q$, we can see how close the protocol is to perfect fidelity
\be 
|\mK|_{\max} \app 1-\f{\pi^2}{4 q^2},\; \f{\max(I_{RT})}{2\log 2}\app1-\f{\pi^2}{4\log 2}\cdot \f{\log q}{q^2}
\ee

For high temperatures, the effect of using multiple fermions  is similarly a replacement  $q \ra q/p$ in the previous result \eqref{eq:highT}. This leads to a maximal value of $|\mK|$ given by
\be\label{eq:maxKh}
|\mK |_\max = \left[\sin \f {\pi} {q/p+2} \right]^{1+2p/q}
\ee
which approaches to 1 as $p\ra \infty$. We expand it at large $p$ and find
\be 
|\mK|_{\max} \app 1-\f{\pi^2 q}{16 p},\; \frac{\max(I_{RT})}{2\log 2}\app 1-\f{\pi^2}{32\log 2}\cdot \f{q\log p/q}{p}
\ee

With this improved protocol, we can encode our qubit into the composite fermion field of $q/2+1$ Majorana fermions in the  low temperature regime or into $p$ Majorana fermions with $p\gg q$ in the high temperature regime  to achieve almost perfect teleportation. The first case has a dual interpretation of a particle passing through the traversable wormhole. The high temperature limit of the SYK model must have a very stringy bulk, and it would be interesting to understand the interpretation of our result in that regime. 

\section{Discussion}\label{sec:discuss}

\subsection{Interference regime} \label{sec:inter}
When an excitation is sent into the wormhole at very early times, its collision with the negative energy squeezed state is highly trans-Planckian and the backreaction is large. 
It was shown in \cite{maldacena2017diving, gao2018regenesis} that nevertheless the left-right correlation function $\mK$ remains nonzero due to an interference effect. We now examine whether this interference effect can lead to teleportation with our protocol. Note that the above large $q$ solution is related to semiclassical gravity plus ``stringy" effects and does not include the interference regime, which goes beyond the leading $1/N$ contributions. Therefore, we have to discuss it separately.

\begin{figure}
\centering
\includegraphics[width=7cm]{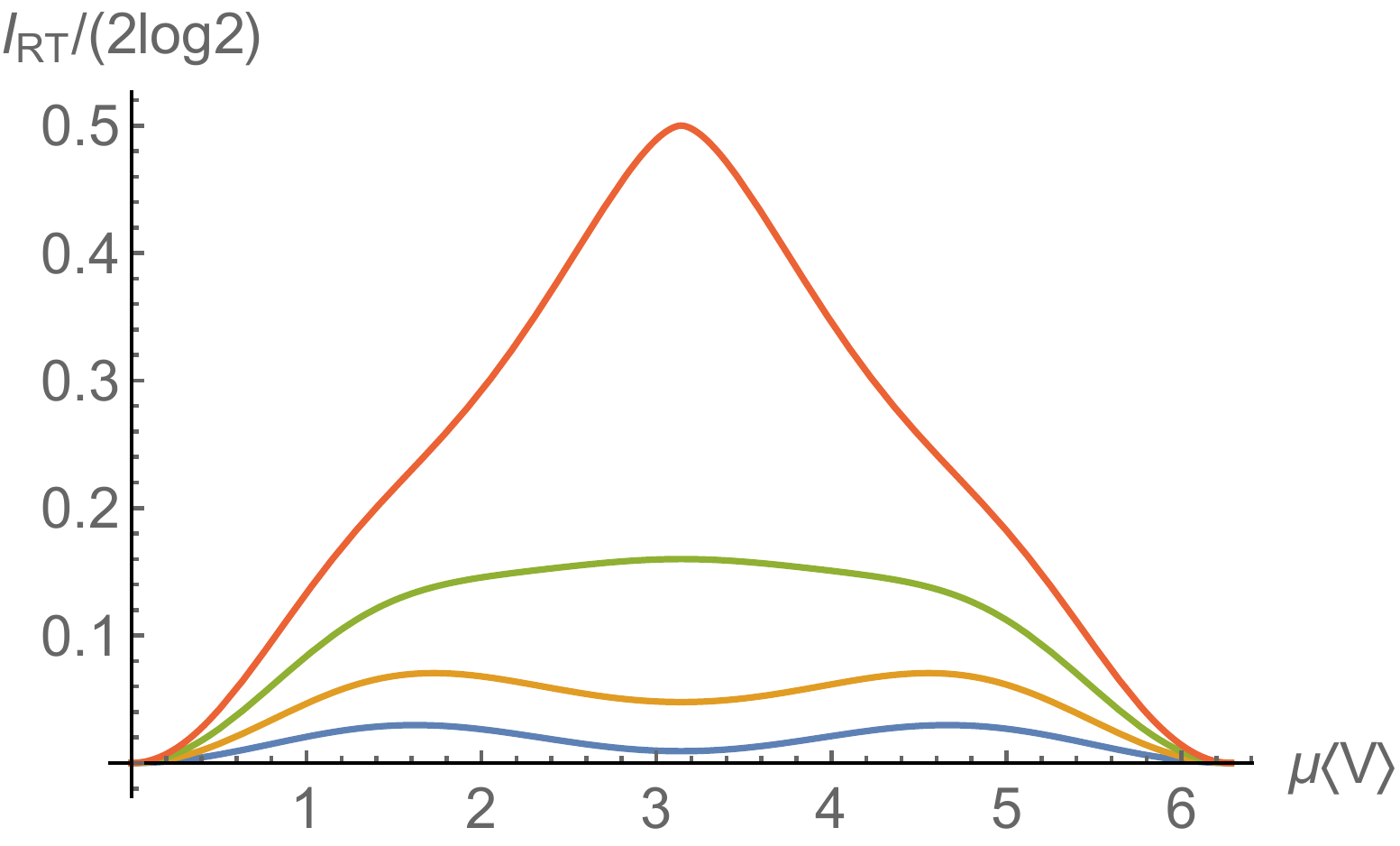}
\caption{Mutual information $I_{RT}$ as a function of $\mu\avg{V}$ for different $|\avg{\psi_{l}^{1}\psi_{r}^{1}}|$. Blue, yellow, green and red represent for $|\avg{\psi_{l}^{1}\psi_{r}^{1}}|=0.2,0.3,0.4,0.5$ respectively. \label{fig:inter}}
\end{figure}

The interference effect relies on the facts that time-ordered correlation functions factorize and out-of-time-ordered correlation functions vanish due to the nature of quantum chaos. Basically, we have 
\begin{equation}
\begin{aligned}\label{eq:5.1}
e^{i\mu V}\bra{\text{tfd}}&\app e^{i\mu\avg V}\bra{\text{tfd}},\\
e^{i\mu V}\psi^{j_1}_{l,r}\cdots \psi^{j_s}_{l,r}\bra{\text{tfd}}&\app\psi^{j_1}_{l,r}\cdots \psi^{j_s}_{l,r}\bra{\text{tfd}}
\end{aligned}
\end{equation}
where $\avg{V}=i\avg{\psi_l(0)\psi_r(0)}/q$ is an order 1 number. Note that the large $q$ solution above does not capture the interference regime because $\mK$ vanishes for large $t$ and $t'$ in \eqref{eq:K} whereas the interference effect implies that it should be $2\Re(e^{i\mu\avg{V}}\avg{\psi_l(-t')\psi_r(t)})$.
Using \eqref{eq:5.1} we find that
\begin{align}
\avg{\{\psi_{l}^{1},[\psi_{l}^{2},\mU^\dagger\psi_{r}^{1}\psi_{r}^{2}\mU]\}} & \ra 4\sin^2\f{\mu\avg V}{2}\avg{\psi_{l}^{1}\psi_{r}^{1}}^{2}\\
\avg{\{\psi_{l}^{1},\mU^\dagger\psi_{r}^{1}\mU\}} & \ra2i\sin\mu\avg V\avg{\psi_{l}^{1}\psi_{r}^{1}}\\
\avg{\{\psi_{l}^{2},\psi_{l}^{1}\mU^\dagger\psi_{r}^{2}\mU\psi_{l}^{1}\}} & \ra0
\end{align}
which leads to 
\begin{align}
\r_{11} & =\f 12-2\sin^2\f{\mu\avg V}{2}\avg{\psi_{l}^{1}\psi_{r}^{1}}^{2}\\
|\r_{14}| & =\sin\mu\avg V\left|\avg{\psi_{l}^{1}\psi_{r}^{1}}\right|
\end{align}
where $\avg{\cdot}$ is short for $\avg{\text{tfd}|\cdot|\text{tfd}}$. Note that $\avg{\psi_{l}^{1}\psi_{r}^{1}}$ is pure imaginary because the correlation function  between $\psi_l$ and $i\psi_r$ is real. Moreover,  $|\avg{\psi_{l}^{1}\psi_{r}^{1}}|$ is bounded by $1/2$ via Schwarzian inequality. For different values of $|\avg{\psi_{l}^{1}\psi_{r}^{1}}|$, we plot $I_{RT}$ as a function of $\mu\avg{V}$ in Fig. \ref{fig:inter}. 

In the low temperature limit, $|\avg{\psi_{l}^{1}\psi_{r}^{1}}|\sim\avg{V}$ scales as $\f 1 2(\pi/\b\mJ)^{2/q}$ which is a small number (much smaller than order 1). From Fig. \ref{fig:inter} we see that $I_{RT}$ is quite small and not sufficient for successful teleportation. More precisely, one can show that 
\be
I_{RT}\sim (\b\mJ)^{-8/q}
\ee
On the other hand, in the high temperature limit, $\avg{\psi_{l}^{1}\psi_{r}^{1}}\ra i/2$, which leads to an order 1 magnitude for $I_{RT}$. In this case, $I_{RT}$ is bounded by $\log 2$. Although the mutual information does not reach maximality, for appropriate choices of $\mu$ some quantum information is teleported via the interference effect, as we will discuss in Sec. \ref{sec:classical}.

\subsection{Partial coupling} \label{sec:partial}

In this section we consider a simple generalization of the protocol in which only a fraction of the Majorana fermions are coupled in the interaction Hamiltonian. This is also required for the coupling being interpreted as a classical channel as discussed in Sec. \ref{sec:setup}. It is particularly interesting to consider measuring the correlation function of fermions
that do not appear in the interaction. The quantum chaotic nature of the SYK Hamiltonian implies that there need be no relation between the fermions that are coupled with the fermions that are transmitted, as expected in the gravity description. 

We consider coupling only the first $\a N$  Majorana fermions\footnote{Here we assume $\a N$ is an integer of order $N$.}
\begin{equation}
H_{int}=-\mu\tilde{V}=-\f{i\mu}{q\a N}\sum_{j=1}^{\a N}\psi_{l}^{j}\psi_{r}^{j}
\end{equation}
and we calculate the correlation function for the last $(1-\a) N$ Majorana
fermions 
\begin{equation}
\sum_j {}^\prime\avg{I|e^{-\b H/2}\bar{\mT}[e^{-\mu\tilde{V}(\b/4+\e)}e^{\mu\tilde{V}(\b/4-\e)}\psi^{j}(\tau_{a})\psi^{j}(\tau_{b})]|I}
\end{equation}
where $\sum'$ means summing $j$ from $\a N+1$ to $N$. It is clear that  $\psi^{j}$ can freely move across $\tilde{V}$ for $j=\a N+1,\cdots,N$, so they do not obey a twisted boundary condition.
One might naively conclude that such correlation functions do not have
any of the nontrivial features of $\mG_{\mu}$ in the various subregions in
Fig. \ref{fig:The-function-} that we found before, in particular that there would be no exponential growth in subregion
$c$ after continuation to Lorentz signature. 

However, this expectation
is incorrect. Indeed, from the general analysis of the SYK model, the low
energy effective action is the Schwarzian that universally couples to  all fermions \cite{maldacena2016remarks}. Thus the
OTOC between different species of fermions should  have
exponential growth at order $1/N$. Moreover, traversable wormholes should exist for generic coupled  operators based on analysis of both the gravity \cite{gao2017traversable} and CFT \cite{gao2018regenesis} sides. 

The resolution is that we need to treat the effective action
(\ref{eq:eff-act}) more carefully. As we separated the fermions into two
groups, we need to define two different $\S$ and $G$ fields
\begin{align}
G_{1}(\tau,\tau')&=\f 1 {\a N}\sum_{i=1}^{\a N}\psi^{i}(\tau)\psi^{i}(\tau')\\
G_{2}(\tau,\tau')&=\f 1{(1-\a)N}\sum_{i=\a N+1}^{N}\psi^{i}(\tau)\psi^{i}(\tau')
\end{align}
The instantaneous coupling adds new terms to the action
\begin{align}
\d S= & \f{\mu}{2q}\left[G_{1}(3\b/4+\e,\b/4-\e)-G_{1}(\b/4-\e,3\b/4+\e)\right.\nonumber \\
 & \left.+G_{1}(\b/4+\e,3\b/4-\e)-G_{1}(3\b/4-\e,\b/4+\e)\right]
\end{align}
Therefore the effective action becomes
\begin{align} \label{eq:5.12S}
S=\f {N}{2} \left(\log\f{\det(\del_{\tau}-\S_{2})]^{\a-1}}{\det(\del_{\tau}-\S_{1})^{\a}}+\iint\left[\a\S_{1}G_{1}+(1-\a)\S_{2}G_{2}-\f {J^{2}} {q}(\a G_{1}+(1-\a)G_{2})^{q}\right]\right)+\d S
\end{align}
which leads to the following equation of motion
\begin{align}
G_{i} =&[\del_{\tau}-\S_{i}]^{-1},\\
\S_{1}=&J^{2}(\a G_{1}+(1-\a)G_{2})^{q-1}+\f{\mu}{q\a N}\tilde{\d},\\
\S_{2}=&J^{2}(\a G_{1}+(1-\a)G_{2})^{q-1},\\
\tilde{\d} =&\d_{\tau,\b/4-\e}\d_{\tau',3\b/4+\e}-\d_{\tau,3\b/4+\e}\d_{\tau',\b/4-\e}-\d_{\tau,\b/4+\e}\d_{\tau',3\b/4-\e}+\d_{\tau,3\b/4-\e}\d_{\tau',\b/4+\e}
\end{align}
where $i=1,2$ and $\d_{a,b}$ is short for $\d(a-b)$. As expected $\S_{2}$ and
$G_{2}$ does not directly have a delta function in their equations
of motion. However, the two parts interact with each other and the
twist boundary condition for $G_{1}$ affect $G_{2}$ accordingly. As discussed in Sec. \ref{sec:syk}, the equations of motion only captures the linearized version of the twist boundary condition. Luckily, in the large $q$ solution, this is enough. To be precise, let us expand
\begin{equation}
[G_{i}]\app [G_{0}]+[G_{0}][\S_{i}][G_{0}]
\end{equation}
Defining $G_{+}=\a G_{1}+(1-\a) G_{2}$ and $G_-=G_1-G_2$, the equations of motion become
\begin{align}
[G_{+}]=&[G_{0}]+\f{\mJ^{2}}q [G_{0}][(2G_{+})^{q-1}][G_{0}]+\f{\mu}{qN}[G_{0}][\tilde{\d}][G_{0}]\\
[G_{-}] =&\f{\mu}{q\a N}[G_{0}][\tilde{\d}][G_{0}]
\end{align}
The first equation is exactly the same as the one for $\mG_{\mu}$ in our earlier analysis, and we can solve it with the same ansatz
\begin{equation}
G_{+}=G_{0}e^{\s_{+}/(q-1)}
\end{equation}
where $\s_{+}$ obeys the boundary conditions (\ref{eq:bd-1})-(\ref{eq:bd-3}). The second equation can be evaluated explicitly
\begin{align}
G_{-}= & \f{\mu}{q\a N}\sum_\pm[\pm G_{0}(\tau,\b/4\mp\e)G_{0}(3\b/4\pm\e,\tau')\mp G_{0}(\tau,3\b/4\pm\e)G_{0}(\b/4\mp\e,\tau')]
\end{align}
In the fundamental region, the configuration is very simple
\begin{equation}
G_{-}=-\f{\mu}{2q\a N}\;\text{in \ensuremath{c}},\;G_{-}=0\;\text{in others}
\end{equation}
Since $G_{1}=G_{+}+(1-\a)G_{-}$ and $G_{2}=G_{+}-\a G_{-}$,
they are identical in all subregions of fundamental region except 
$c$. In subregion $c$, both $G_{1}$ and $G_{2}$ have exponential
growth after continuation to Lorentz signature as $G_{+}$ does. However, on the boundary
of $c$, the boundary condition for $\s_{+}$ is
\begin{equation}
\s_{+}(\text{inside }c)=\s_{+}(\text{outside }c)-\f{\mu}{N}
\end{equation}
To leading order in $1/q$, this shift perfectly cancels the constant
shift in $G_{-}$,
\begin{align}
G_{2}(\text{inside }c)&\app \f 12(1+\f 1q\s_{+}(\text{inside }c))+\f{\mu}{2qN}\nn\\
&=\f 12(1+\f 1q\s_{+}(\text{outside }c))\app G_{2}(\text{outside }c)
\end{align}
which shows that $G_{2}$ is continuous. This analysis clearly shows
the crucial feature of traversable wormhole teleportation that the
coupled qubits between two systems can be chosen independently from
the teleported qubits.

Note that the number of coupled qubits must be of order $N$ %, namely $\a$ in Sec. \ref{sec:partial} being order 1, 
to guarantee that the path integral over the effective action \eqref{eq:5.12S} is dominated by the classical solution. Otherwise, one must treat all $\a N\sim O(1)$ Majorana Fermions with full quantum corrections, which is beyond the regime of validity of the calculations in this paper. 

\subsection{Teleportation of multiple qubits} \label{sec:compare}

If we want to teleport $n$ qubits together, we can simply prepare $R=\cup_i R_i, T=\cup_i T_i$ and $Q=\cup_i Q_i$ for $i=1,\cdots,n$ and apply $n$ SWAP operators associated to $n$ different (Dirac) fermions in the SYK model. As these SWAP operators have different fermion indices, they do not talk to each other, due to the $SO(N)$ approximate symmetry. Therefore, correlation functions with twist boundary conditions factorize into independent channels in the large $N$ limit. It follows that the reduced density matrix between $R$ and $T$ is tensor product of $n$ identical matrices, each of which is given as before \eqref{eq:17}. The caveat in this argument is that the number of teleported qubit $n$ must be $o(N)$, otherwise the factorization of the correlation functions at large $N$ breaks down. Since $o(N)<O(N)$, this is consistent with the teleportation information bound: 2 classical bits can only teleport at most one qubit. This is also consistent with the upper bound of total number of teleported qubit from gravitational computation \cite{Freivogel:2019whb}.

Although the traversable wormhole teleportation does not teleport the maximal number of qubits, it does have an advantage that the qubits are strongly error protected \cite{almheiri2015bulk}. Indeed, as we can choose arbitrary $O(N)$ Majorana fermions in the interaction $V$, it does not matter what happens to other uncoupled ones. In the quantum information language, when some qubits are corrupted or erased, we can still measure other intact qubits in the $l$ system to complete the teleportation protocol. Holographically, the dual picture is a qubit traveling through a traversable wormhole. As the location of qubit is in the deep interior, it is not sensitive to local boundary changes, which correspond to simple ($O(1)$ qubit) errors.

\subsection{Classical simulation and separability} \label{sec:classical}

As we have seen in Sec. \ref{sec:improve}, an appropriate choice of SWAP operator leads to almost perfect fidelity in the large $q$ limit. However, in the interference regime and for other choices of the SWAP in the large $q$ solution, the mutual information can still be $O(1)$ although submaximal, so there is not perfect teleportation. Here we examine in more detail whether teleportation in those cases really communicates  quantum information that cannot be simulated by a purely classical communication protocol. In other words, we need to investigate if $\r_{RT}$ is separable.

Generally, a density matrix on two systems $\r_{RT}$ is separable if it can be written in the form of sum of tensor products of density matrices in each subsystem, 
\be 
\r_{RT}=\sum_i p_i \r_R^i \otimes \r_T^i,\; \sum_i p_i =1,\; p_i>0,\; \forall i
\ee
Here $p_i$ can be understood as the classical probability for each tensor product state. By definition, it is clear that the mutual information is purely classical if the density matrix $\r_{RT}$ is separable. In such a case, we could achieve the same result with classical simulation, and our quantum teleportation protocol would have failed.

For a general bipartite density matrix, determining if it is separable is an NP-hard problem \cite{gurvits2003classical, gharibian2008strong}. However, for our pair of single qubit systems, the sufficient and necessary condition is the Peres-Horodecki criterion \cite{peres1996separability, horodecki1997separability}, which states that if the partial transposed density matrix has a negative eigenvalue, then the density matrix is quantum entangled. By \eqref{eq:46-0}-\eqref{eq:46}, the partial transpose of the density matrix $\r^{\text{T}_R}$ is simply obtained  by exchanging $\r_{14}$ with $\r_{23}$. 

In our large $q$ solution, the eigenvalues of $\r^{\text{T}_R}$ are
\be 
\f 1 4 (1+\mK^2) ~(\text{multiplicity 2}),\quad \f 1 4 (1 \pm 2 \mK - \mK^2)
\ee
The first expression is always positive and the second one could be negative when $|\mK|>\sqrt{2}-1 \app 0.414$. For low temperatures with the improved protocol, the maximum value of $|\mK|$ is given by  \eqref{eq:maxKp}, which satisfies this bound when $0.094q<p<1.260p$.  For high temperatures with the improved protocol, the maximum value of $|\mK|$ is given by  \eqref{eq:maxKh}, which satisfies the bound when $p> 0.336 q$. In those cases, R and T have some quantum entanglement.

In the interference regime, the eigenvalues of $\r^{\text{T}_R}$ are
\begin{align}
\f 1 4 (1 + 2 U^2 - 2 U^2 \cos\mu \avg{V}) ~(\text{multiplicity 2}),\\
\f 1 4 (1 - 2 U^2 + 2 U^2 \cos\mu \avg{V} \pm 2 U \sin\mu \avg{V}),
\end{align}
where $U\equiv -i\avg{\psi_l(-t') \psi_r(t)}\in (0,1/2)$. The first expression is always positive and the second one could be negative when
\be  
\f 1 2 \geq U > \f {-\sin\mu \avg{V}+\sqrt{(1-\cos\mu \avg{V})(3+2\cos\mu \avg{V})}}{2(1-\cos\mu \avg{V})}
\ee
It is clear that in the high temperature limit, $U=|V|=1/2$ and above inequality is satisfied when $2.294+4k\pi<\mu<7.131+4k\pi$ for $k\in\Z$. In this case, there is teleportation of some quantum information in the interference regime.

\section{Conclusion}\label{sec:conclusion}

In this paper, we proposed a simple teleportation protocol using a pair of SYK systems. The protocol requires the
two systems prepared in the highly entangled thermofield double state. There are three steps:
\begin{enumerate}
  \item Apply a SWAP operation in one system to insert a qubit at early (negative) time $t=-t_{in}$;
  \item Evolve the state to $t=0$ and apply a weak coupling between two SYKs;
  \item Evolve the state to $t\app t_{in}$ and apply a SWAP to extract the qubit.
\end{enumerate}

We calculated the mutual information $I_{RT}$ between a reference system and the final qubit to confirm the effectiveness of this protocol. The protocol has a clear holographic interpretation in which the teleported qubit goes through a traversable wormhole in a nearly AdS$_2$ spacetime. At low temperature, the result matches the semiclassical gravity analysis in two aspects: first, we see a high peak in $I_{RT}$ at a specific time that is related to the geodesics through the traversable wormhole; second, there is a range of time in which time ordering of the signal is preserved. We take these as signs that the information travels smoothly through the throat of a traversable wormhole. As the temperature increases, we identify some ``stringy" corrections to correlation functions that will soften the peak in $I_{RT}$.

Our protocol is quite similar to standard teleportation in the sense that the weak coupling between
two SYKs can be implemented as a measurement in one system and a corresponding unitary in the other.
However, the measurement in our protocol is in the computational basis thanks to scrambling, and
the teleported qubit is highly error protected. This protocol allows teleportation of $o(N)$ qubits simultaneously, at the cost of communication of $O(N)$ bits of classical information, which obeys the information bound on quantum teleportation.

The protocol can be improved to approach perfect fidelity by an appropriate choice of SWAP operator. For
large $q$ SYK at low temperatures, using a SWAP of the qubit into $q/2 + 1$ Majorana fermions, we find that the mutual information $I_{RT}$ tends to 1 when $q\ra\infty$. Similarly, at high temperature, SWAP of the qubit
into $p$ Majorana fermions with $p \gg q$ also results in mutual information close to 1.

At very late times, we also analyzed the teleportation effect in the fully scrambled interference regime. It turns out that the
interference effect makes teleportation possible only at high temperature rather than low temperature with
our simple protocol. This corresponds to very late time black holes, for which Kitaev and Yoshida proposed an algorithm \cite{yoshida2017efficient} involving complicated operators to decode the scrambled qubit. Our protocol has a much simpler structure and works for both around the scrambling
time at all temperatures, and for very late times at high temperature.

\begin{acknowledgments}
We thank Xun Gao, Yingfei Gu, Emil Khabiboulline, Thomas Schuster, Maria Spiropulu and Norman Yao for helpful and stimulating discussions. PG would like to thank Tsung-Dao Lee Institute of Shanghai Jiaotong University for hospitality during the period of this project. PG and DLJ are both supported by DOE award DE-SC0019219. Besides, PG is supported by the US Department of Energy grants DE-SC0018944 and DE-SC0019127, and also the Simons foundation as a member of the {\it It from Qubit} collaboration.
\end{acknowledgments}

\appendix

\section{Solution of $\s$} \label{app}
In this appendix, we provide the details of how to determine $\s$. Let us assume that
\begin{equation}
g_{0}^{\mu}=\widetilde{SL}_{\mu}(f_{0}^{-\mu})\equiv\f{\tilde{a}_{\mu}+\tilde{b}_{\mu}f_{0}^{-\mu}}{\tilde{c}_{\mu}+\tilde{d}_{\mu}f_{0}^{-\mu}}
\end{equation}
which leads to
\begin{align}
(f_{0}^{\mu},\widetilde{SL}_{\mu}(f_{0}^{-\mu}))&\simeq(\widetilde{SL}_{-\mu}(f_{0}^{\mu}),f_{0}^{-\mu})\simeq(f_{0}^{\mu},\widetilde{SL}_{-\mu}^{-1}(f_{0}^{-\mu})),\nn\\\widetilde{SL}_{-\mu}^{-1}(f)&\equiv\f{\tilde{d}_{-\mu}+\tilde{b}_{-\mu}\mJ^{2}f}{\mJ^{2}(\tilde{c}_{-\mu}+\tilde{a}_{-\mu}\mJ^{2}f)}\label{eq:122}
\end{align}
where the first $\simeq$ is due to the symmetry (\ref{eq:119}) and the
second $\simeq$ is due to the global $SL(2)$ symmetry (\ref{eq:global-sl2})
of the large $q$ solution. When $\mu=0$, we see from (\ref{eq:mu0-sln})
that $\widetilde{SL}$ is the identity, from which we can fix $\tilde{b}_{\mu}\tilde{c}_{\mu}-\tilde{a}_{\mu}\tilde{d}_{\mu}=1$
because we want the solution to smoothly reduce to the $\mu=0$ case.
It follows that
\begin{align}
&\f{\tilde{a}_{\mu}+\tilde{b}_{\mu}f_{0}^{-\mu}}{\tilde{c}_{\mu}+\tilde{d}_{\mu}f_{0}^{-\mu}}=\f{\tilde{d}_{-\mu}+\tilde{b}_{-\mu}\mJ^{2}f_{0}^{-\mu}}{\mJ^{2}(\tilde{c}_{-\mu}+\tilde{a}_{-\mu}\mJ^{2}f_{0}^{-\mu})}\nn\\
&\implies\tilde{a}_{\mu}\mJ^{2}-\tilde{d}_{-\mu}=\tilde{b}_{\mu}-\tilde{b}_{-\mu}=\tilde{c}_{\mu}-\tilde{c}_{-\mu}=0\label{eq:123}
\end{align}
Similarly, $f_{0}^{\mu}$ must be related to $f_{0*}^{\mu}$ by an
$SL(2)$ transformation, explicitly 
\begin{equation}
f_{0}^{\mu}=sl_{\mu}(f_{0*}^{\mu})\equiv\f{\a_{\mu}+\b_{\mu}f_{0*}^{\mu}}{\g_{\mu}+\d_{\mu}f_{0*}^{\mu}}
\end{equation}
This $SL(2)$ squares to the identity because $f_{0}^{\mu}=sl_{\mu}(f_{0*}^{\mu})=sl_{\mu}(sl_{\mu}(f_{0}^{\mu}))$,
which implies that
\begin{equation}
\b_{\mu}+\g_{\mu}=0,\;\text{or }\a_{\mu}=0,\d_{\mu}=0\label{eq:124}
\end{equation}
Comparing with (\ref{eq:mu0-sln}), we see that
\begin{equation}
f_{*}=\f{-\f 1{\mJ}\sin2v+\cos2v\cdot f}{-\cos2v-\mJ\sin2v\cdot f},~f=\f 1{\mJ}\tan(\w\tau+v)
\end{equation}
which has nonzero $\a_{\mu}$ and $\d_{\mu}$ under $\tau$ reflection.
Therefore we must take the former choice in (\ref{eq:124}) and fix
$\b_{\mu}^{2}+\a_{\mu}\d_{\mu}=1$. 

Since we have two $SL(2)$ symmetries in region $b$, they need 
to be consistent with each other
\begin{equation}
(f_{0}^{\mu},\widetilde{SL}_{\mu}(f_{0}^{-\mu}))\simeq(f_{0*}^{\mu},\widetilde{SL}_{\mu}(f_{0*}^{-\mu}))
\end{equation}
In other words, as $sl_{\mu}$ transforms $f_{0*}^{\mu}$ to $f_{0}^{\mu}$,
it must transform $\widetilde{SL}_{\mu}(f_{0*}^{-\mu})$ to $\widetilde{SL}_{\mu}(f_{0}^{-\mu})$
accordingly. We can follow similarly to the derivation of (\ref{eq:122})
to get
\begin{equation}
sl_{\mu}^{-1}(\widetilde{SL}_{\mu}(f))=\widetilde{SL}_{\mu}(sl_{-\mu}(f))
\end{equation}
which leads to
\begin{align}
\tilde{a}_{\mu}\b_{\mu}+\tilde{c}_{\mu}\d_{\mu}/\mJ^{2} & =\tilde{b}_{\mu}\a_{-\mu}-\tilde{a}_{\mu}\b_{-\mu},\\
\tilde{b}_{\mu}\b_{\mu}+\tilde{d}_{\mu}\d_{\mu}/\mJ^{2}&=\tilde{a}_{\mu}\d_{-\mu}+\tilde{b}_{\mu}\b_{-\mu}\\
\mJ^{2}\tilde{a}_{\mu}\a_{\mu}-\tilde{c}_{\mu}\b_{\mu} & =\tilde{d}_{\mu}\a_{-\mu}-\tilde{c}_{\mu}\b_{-\mu},\\
\mJ^{2}\tilde{b}_{\mu}\a_{\mu}-\tilde{d}_{\mu}\b_{\mu}&=\tilde{c}_{\mu}\d_{-\mu}+\tilde{d}_{\mu}\b_{-\mu}
\end{align}
which can be solved as
\begin{equation}
\begin{aligned}
\tilde{a}_{\mu}&=\f{-\tilde{b}_{\mu}\b_{-\mu}\pm\sqrt{\tilde{b}_{\mu}^{2}-\d_{\mu}\d_{-\mu}/\mJ^{2}}}{\d_{-\mu}},\\
\tilde{c}_{\mu}&=\mJ^{2}\f{\tilde{b}_{\mu}(1+\b_{\mu}\b_{-\mu})\mp(\b_{\mu}+\b_{-\mu})\sqrt{\tilde{b}_{\mu}^{2}-\d_{\mu}\d_{-\mu}/\mJ^{2}}}{\d_{\mu}\d_{-\mu}}\label{eq:130-1}
\end{aligned}
\end{equation}
with $\tilde{b}_{\mu}=\tilde{b}_{-\mu}$, $\tilde{d}_{\mu}=\mJ^{2}\tilde{a}_{-\mu}$
and $\a_{\mu}=(1-\b_{\mu}^{2})/\d_{\mu}$ by definition. These are
all constraints resulting from the symmetries and smoothness requirements.

We can apply a similar analysis to regions $a$ and $e$. There is only
one symmetry in each of these two regions, so the constraint is much
weaker. Taking all these into account, we assign functions to each
subregion in Fig. \ref{fig:The-function-} with only three functions
$f^{\mu}$, $g_{1}^{\mu}$ and $g^{\mu}$ (here we simplify notation
$f_{0}^{\mu}\ra f^{\mu}$), and four $SL(2)$ transformations $\widetilde{SL}_{\mu}$,
$\widehat{SL}_{\mu}$, $\overline{SL}_{\mu}$ and $sl_{\mu}$ to be
determined, where $\widehat{SL}_{\mu}$ and $\overline{SL}_{\mu}$
obey
\begin{align}
\hat{a}_{\mu}\mJ^{2}-\hat{d}_{-\mu}=\hat{b}_{\mu}-\hat{b}_{-\mu}=\hat{c}_{\mu}-\hat{c}_{-\mu}=\bar{a}_{\mu}\mJ^{2}-\bar{d}_{-\mu}=\bar{b}_{\mu}-\bar{b}_{-\mu}=\bar{c}_{\mu}-\bar{c}_{-\mu}=0\label{eq:131}
\end{align}
with $\hat{b}_{\mu}\hat{c}_{\mu}-\hat{a}_{\mu}\hat{d}_{\mu}=1$ and
$\bar{b}_{\mu}\bar{c}_{\mu}-\bar{a}_{\mu}\bar{d}_{\mu}=-1$. Here
the normalization for $\overline{SL}_{\mu}$ is $-1$ to agree in the $\mu=0$ case with $g_{1}^{\mu}=\overline{SL}_{\mu}(g_{1*}^{-\mu})=f$
in (\ref{eq:mu0-sln}). 

The twist boundary conditions on the four blue segments in Fig. \ref{fig:The-function-}
are given by \eqref{eq:b1}-\eqref{eq:b4}. We can use these equations to determine
both $\overline{SL}_{\mu}$ and $\widehat{SL}_{\mu}$ based on the
value of $f^{\mu}$ and $g_{1}^{\mu}$ on the boundaries. For notational
simplicity, we define $f_{\pm}^{\mu}\equiv f^{\mu}(\tau_{\pm})$, $f_{\pm*}^{\mu}\equiv f^{\mu}(-\tau_{\pm})$
and similarly for $g_{1}^{\mu}$ and their derivatives. Using (\ref{eq:130-1})
and $f_{-}^{\mu}$ we can determine $g_{1-}^{\mu}$ and $g_{1-}^{\mu}{'}$
via (\ref{eq:b2}). Based on these values, we can determine $\widehat{SL}_{\mu}$
completely via (\ref{eq:b3}), the constraints (\ref{eq:131}) and the value
(\ref{eq:130-1}). It turns out that
\begin{equation}\label{eq:a13}
\hat{b}_{\mu}=\mp\tilde{c}_{\mu},\;\hat{c}_{\mu}=\mp\tilde{b}_{\mu},\;\hat{a}_{\mu}=\pm\tilde{a}_{-\mu},\;\hat{d}_{\mu}=\mJ^{2}\hat{a}_{-\mu}
\end{equation}
As both signs give the same $\widehat{SL}_{\mu}$, we will choose
the lower sign without loss of generality. To solve for $\overline{SL}_{\mu}$ is similar. The only
difference is that it involves $f_{-*}^{\mu}$ that is related to
$f_{-}^{\mu}$ via $sl_{\mu}$. It turns out that one can determine $\bar{a}_{\mu}$,
$\bar{b}_{\mu}$, $\bar{c}_{\mu}$ and $\bar{d}_{\mu}$ explicitly
in terms of $\tilde{b}_{\mu}$, $\b_{\pm\mu}$, $\d_{\pm\mu}$ and
$f_{\pm}^{\mu}$ only. However, the expression is very complicated
and we do not reproduce it here. At this point, the free parameters have been
reduced to $\tilde{b}_{\mu},\b_{\pm\mu},\d_{\pm\mu}$.

As argued in Sec. \ref{sec:large-q}, there are still too many degree of freedom to uniquely fix the solution. Taking the ansatz \eqref{eq:139}, we immediately get
\begin{align}
\b_{\mu}=\b=\cos2v,\;\d_{\mu}=\d=-\mJ\sin2v\implies\text{\ensuremath{\tilde{a}}}_{\mu}=\tilde{d}_{\mu}=\hat{a}_{\mu}=\hat{d}_{\mu}=0,\;\tilde{c}_{\mu}=\hat{b}_{\mu}=1
\end{align}
On the other hand, we know
\begin{equation}
(\overline{SL}_{\mu}(g_{1*}^{-\mu}),g_{1}^{\mu})\simeq\left(\f{a+bf}{c+df},f\right),\quad bc-ad=1\label{eq:141}
\end{equation}
for some $a,b,c,d$. Applying this to the formula for $e^{\s}$ and checking the UV condition,
we find 
\begin{equation}
\begin{aligned}
b=c=\cos\p,\;d=-a&\mJ^{2}=-\mJ\sin\p,\\
e^{\s}=\f{\w^{2}}{\mJ^{2}\cos^{2}(\w\tau_{ab}+\p)},&\;\w=\mJ\cos(\p-\w\b/2)\label{eq:142}
\end{aligned}
\end{equation}
Now we can forget about $\overline{SL}_{\mu}$ and directly impose
the symmetry conditions: $\mu\ra-\mu$, $\tau_{a,b}\ra-\tau_{b,a}$.
This requires that $\p$ is invariant under $\mu\ra-\mu$. In order
to more simply satisfy the twist boundary condition, we use the global $SL(2)$ to
write
\begin{equation}
\left(\f{a+bf}{c+df},f\right)\simeq\left(F\left(\f{a+bf}{c+df}\right),G(f)\right)
\end{equation}
where $F$ is given by (\ref{eq:global-sl2})
\begin{equation}
F(f)=\f{D_{\mu}-C_{\mu}\mJ^{2}f}{\mJ^{2}(-B_{\mu}+A_{\mu}\mJ^{2}f)}\label{eq:144-1}
\end{equation}
Let us check twist boundary conditions using (\ref{eq:139}), (\ref{eq:144-1})
and (\ref{eq:142}):
\begin{align}
G(f_{-})&=f_{-},\;G'(f_{-})=e^{-\mu/N},\\
F\left(\f{a+bf_{-*}}{c+df_{-*}}\right)&=f_{-*},\;\f {F'\left(\f{a+bf_{-*}}{c+df_{-*}}\right)}{(c+df_{-*})^{2}}=e^{\mu/N}
\end{align}
These four equations determine the four variables: $A_{\mu},B_{\mu},C_{\mu},\p$,
in which
\begin{equation}
\cot\f{\p}2=\cot2\w\tau_{-}\text{ or }\p=0
\end{equation}
As we would like to match on to the $\mu=0$ solution smoothly, we must choose
the second case. It follows that
\begin{align}
A_{\mu}&=\f{2f_{-}\sinh\f{\mu}{2N}}{1+\mJ^{2}f_{-}f_{-*}},~B_{\mu}=\f{\mJ^{2}f_{-}f_{-*}e^{\mu/(2N)}+e^{-\mu/(2N)}}{1+\mJ^{2}f_{-}f_{-*}},\\
C_{\mu}&=\f{\mJ^{2}f_{-}f_{-*}e^{-\mu/(2N)}+e^{\mu/(2N)}}{1+\mJ^{2}f_{-}f_{-*}},~D_{\mu}=\f{2\mJ^{2}f_{-*}\sinh\f{\mu}{2N}}{1+\mJ^{2}f_{-}f_{-*}}
\end{align}
and UV condition becomes
\begin{equation}
\w=\mJ\cos(\w\b/2)\label{eq:148-2}
\end{equation}

Lastly, we will solve for $g^{\mu}$. This is easier to do in terms of the
unshifted $\tau_{a}$ coordinate. In the unshifted coordinate, 
\begin{equation}
f^{\mu}=\f {\tan(\w(\tau-\b/2)+v)}{\mJ}=\f{-\f 1{\mJ}\tan\w\b/2+f}{1+\mJ\tan\w\b/2\cdot f}\label{eq:149}
\end{equation}
This is an $SL(2)$ transformation. We can simply assume that in the unshifted
coordinate $f^{\mu}=f$, and find the most general $g^{\mu}$ that satisfies the
UV condition. Then doing a corresponding $SL(2)$ transformation
(\ref{eq:global-sl2}) to $g^{\mu}$ will give the solution in the shifted
coordinate. Let us assume $g^{\mu}=H(f)$ for some functional of $f$.
The UV condition becomes
\begin{equation}
\f{H'(f)f'^{2}}{(1+\mJ^{2}H(f)f)^{2}}=1\label{eq:148-1}
\end{equation}
Note that for our special choice of $f$, $f'$ is related to $f$
by
\begin{equation}
f'^{2}=\f{\w^{2}}{\mJ^{2}}\sec^{4}(\w\tau+v)=\f{\w^{2}}{\mJ^{2}}(1+\mJ^{2}f^{2})^{2}
\end{equation}
Plug this back to (\ref{eq:148-1}) and solve for $H$ as
\begin{align}
H(f)=\f 1 {\mJ }\f{h_{\mu}\sin(\w\tau+v-\phi)-e^{2(\w\tau+v)\tan\phi}\sin(\w\tau+v+\phi)}{h_{\mu}\cos(\w\tau+v-\phi)-e^{2(\w\tau+v)\tan\phi}\cos(\w\tau+v+\phi)}
\end{align}
where $\w=\mJ\cos\phi$ and $h_{\mu}$ is an integration constant. Comparing with (\ref{eq:148-2}),
we have $\phi=\w\b/2$. Before shifting it back to the new coordinate
$\tau_{a}\ra\tau_{a}+\b/2$, we apply (\ref{eq:149}) to $f$ and an
associated $SL(2)$ transformation to $g^{\mu}$:
\begin{equation}
g^{\mu}\ra\f{\f 1{\mJ}\tan\w\b/2-g^{\mu}}{-1-\mJ\tan\w\b/2\cdot g^{\mu}}
\end{equation}
which leads to
\begin{equation}
g^{\mu}=\f 1{\mJ}\f{e^{2(\w\tau+v)\tan\w\b/2}\sin(\w\tau+v)-h_{\mu}\sin(\w\tau+v-\w\b)}{e^{2(\w\tau+v)\tan\w\b/2}\cos(\w\tau+v)-h_{\mu}\cos(\w\tau+v-\w\b)}
\end{equation}
Solving twist boundary condition, we find 
\begin{equation}
h_{\mu}=\f{e^{2(\w \tau_{+}+v)\tan\w\b/2}(e^{-\mu/N}-1)\sin\w\tau_{+-}\sin(\w\tau_{-}-v)}{\cos\w\b/2\cos(\w\tau_{-}+v)-\cos2\w\tau_{-}\cos(\w\tau_{+}-\w\b+v)+e^{-\mu/N}\sin\w\tau_{+-}\sin(\w\tau_{-}-v)}
\end{equation}
which vanishes when $\tau_{+}\ra\tau_{-}$. This is a consistency
requirement. It is important that we need a nontrivial $h_{\mu}$ for
a consistent solution to our problem, even though in the end we do not
need it as we will take the $\tau_{+}\ra\tau_{-}$ limit.

\bibliographystyle{JHEP.bst}
\bibliography{WHT.bib}

\end{document}